\pdfoutput=1
\documentclass[twoside,11pt]{article}

\usepackage{jmlr2e} 
\usepackage{picinpar}
\usepackage{graphicx}
\usepackage{latexsym}
\usepackage{amsmath}
\usepackage{amssymb} 
\usepackage{bm}
\usepackage{natbib}
\usepackage[footnotesize]{caption}
\usepackage{color}
\usepackage{enumitem}

\numberwithin{equation}{section}
\numberwithin{figure}{section}

\definecolor{DarkGreen}{rgb}{0.0, 0.5, 0.0}

\def\HKW94{hkw-twlbpea-94}

\def\hkst94{hkst-rlccbsm-94}
\def\OpHa94{op-gbbrlt-94}
\def\BiMa94{bm-rcmce-93} 

\def\CBFHW93{bfhw-opcs-93}
\def\HW94{hw-wl-94}

\def\KW93{kw-epco-94}

\def\Haykin94{h-nncf-94}

\newcommand{\expgrad}{kw-avegulp-97}

\newcommand{\dotpred}{kw-aep-99}

\newcommand{\myspeech}{swrl-cwfeart-03}
\newcommand{\bhatia}{b-ma-97}

\newcommand{\meg}{trw-meg-05}

\newcommand{\qbayes}{w-qbr-05}

\newcommand{\variance}{wk-olvm-06}
\newcommand{\pca}{wk-rpawtbld-06}

\newcommand{\subspace}{w-ws-07}

\newcommand{\N}{\mathbb{N}}
\newcommand{\p}{\boldsymbol{{p}}}

\newcommand{\ww}{\boldsymbol{\omega}}

\renewcommand{\SS}{\boldsymbol{\mathcal{S}}}

\renewcommand{\aa}{{\boldsymbol{\alpha}}}
\renewcommand{\S}{{\boldsymbol{S}}}
\renewcommand{\P}{{\boldsymbol{P}}}
\newcommand{\s}{{\boldsymbol{s}}}
\newcommand{\h}{{\boldsymbol{h}}}
\newcommand{\w}{{\boldsymbol{w}}}
\newcommand{\T}{\boldsymbol{T}}
\newcommand{\W}{\boldsymbol{W}}
\renewcommand{\H}{\boldsymbol{H}}

\newcommand{\RR}{\boldsymbol{R}}
\newcommand{\AS}{{\mathbb{A}}}
\newcommand{\BS}{{\mathbb{B}}}
\newcommand{\MS}{\mathbb{M}}
\newcommand{\YS}{\mathbb{Y}}

\newcommand{\DD}{\boldsymbol{\mathcal{D}}}

\renewcommand{\gg}{{\boldsymbol{\gamma}}}
\newcommand{\uu}{{\boldsymbol{u}}}
\newcommand{\vv}{{\boldsymbol{v}}}
\newcommand{\UU}{\boldsymbol{\mathcal{U}}}
\newcommand{\WW}{{\boldsymbol{\mathcal{W}}}}

\newcommand{\D}{{\boldsymbol{D}}}
\newcommand{\G}{{\boldsymbol{G}}}

\renewcommand{\t}{\boldsymbol{t}}

\newcommand{\LL}{\boldsymbol{\mathcal{L}}}
\renewcommand{\ll}{{\boldsymbol{\lambda}}}
\renewcommand{\l}{{\boldsymbol{l}}}
\renewcommand{\L}{{\boldsymbol{L}}}
\newcommand{\y}{\boldsymbol{y}}

\newcommand{\R}{\mathbb{R}}
\newcommand{\C}{\boldsymbol{C}}

\newcommand{\E}{{\boldsymbol{E}}}
\newcommand{\F}{{\boldsymbol{F}}}

\newcommand{\V}{\mathbb{V}}

\newcommand{\I}{{\boldsymbol{I}}}
\newcommand{\A}{{\boldsymbol{A}}}
\newcommand{\B}{{\boldsymbol{B}}}
\newcommand{\M}{{\boldsymbol{M}}}
\newcommand{\U}{{\boldsymbol{U}}}
\renewcommand{\u}{\boldsymbol{u}}
\renewcommand{\v}{\boldsymbol{v}}
\newcommand{\x}{\boldsymbol{x}}
\renewcommand{\a}{\boldsymbol{a}}
\renewcommand{\b}{\boldsymbol{b}}
\newcommand{\e}{\boldsymbol{e}}
\newcommand{\m}{\boldsymbol{m}}
\renewcommand{\c}{\boldsymbol{c}}
\newcommand{\zero}{\boldsymbol{0}}
\newcommand{\tr}{{\mathrm{tr}}}
\newcommand{\range}{{\mathrm{range}}}
\newcommand{\supp}{{\mathrm{s}}}

\DeclareMathOperator{\expm}{\mathbf{exp}}
\DeclareMathOperator{\logm}{\mathbf{log}}
\DeclareMathOperator{\diag}{diag}

\DeclareMathOperator*{\argmax}{arg\;max}

\DeclareMathOperator*{\arginf}{arg\;inf}

\ShortHeadings{Calculus for Density Matrices}{Warmuth and Kuzmin}
\firstpageno{1}
\title{Bayesian Generalized Probability Calculus for Density 
Matrices\thanks{Supported by NSF grant IIS 0325363. 
Some of this work was done while
visiting National ICT Australia in Canberra.}}

\author{\name Manfred K. Warmuth \email
manfred@cse.ucsc.edu\\
        \name Dima Kuzmin \email dima@cse.ucsc.edu\\
        \addr Computer Science Department\\
        University of California - Santa Cruz\\
\\
        October 10, 2007
}

\editor{}

\begin{document}
\maketitle
\begin{abstract}
One of the main concepts in quantum physics is a density matrix, 
which is a symmetric positive definite matrix of trace one.
Finite probability distributions can be seen as a special case
when the density matrix is restricted to be diagonal.

We develop a probability calculus based on these more
general distributions that includes definitions of joints, 
conditionals and formulas that relate these, including analogs 
of the Theorem of Total Probability and
various Bayes rules for the calculation of posterior density matrices. 
The resulting calculus parallels the familiar
``conventional'' probability
calculus and always retains the latter as a special case when all
matrices are diagonal. We motivate both the
conventional and the generalized Bayes rule with
a minimum relative entropy principle, where the
Kullbach-Leibler version
gives the conventional Bayes rule and 
Umegaki's quantum relative entropy
the new Bayes rule for density matrices.

Whereas the conventional Bayesian methods maintain uncertainty
about which model has the highest data likelihood, 
the generalization maintains
uncertainty about which unit direction has the largest
variance. Surprisingly the bounds also generalize:
as in the conventional setting 
we upper bound the negative log likelihood of the data 
by the negative log likelihood of the MAP estimator.
\end{abstract}



\section{Introduction}
The main notion of a ``mixture state'' used in quantum physics
is a density matrix. States are unit vectors $\u$ ($\lVert\u\rVert_2 = 1$). For the
sake of simplicity we assume in this paper that
the underlying vector space is $\R^n$ (for finite $n$).
Each state $\u$ (unit column vector in $\R^n$) 
is associated with a {\em dyad} $\u\u^\top \in \R^{n\times n}$.
The dyad $\u\u^\top$ may be seen as a one-dimensional
projection matrix which projects any vector onto direction $\u$. 
These dyads are the elementary events of a 
{\em generalized probability space}. 
It is useful to keep the corresponding ``conventional''
probability space in mind, which consists of a finite set of size $n$. 
The $n$ points are the elementary events
and a probability distribution may be seen as a mixture
over the $n$ points, i.e. such a probability distribution
is specified by $n$ real numbers that are bigger than
zero and add to one.
In the generalized case there are infinitely many dyads
even if the dimension $n$ is finite.\footnote{%
The machinery for infinite dimensional vector spaces 
is available. However, in this paper we start with
the simplest finite dimensional setting.} 

\begin{figure*}[t]
\includegraphics[width=\textwidth]{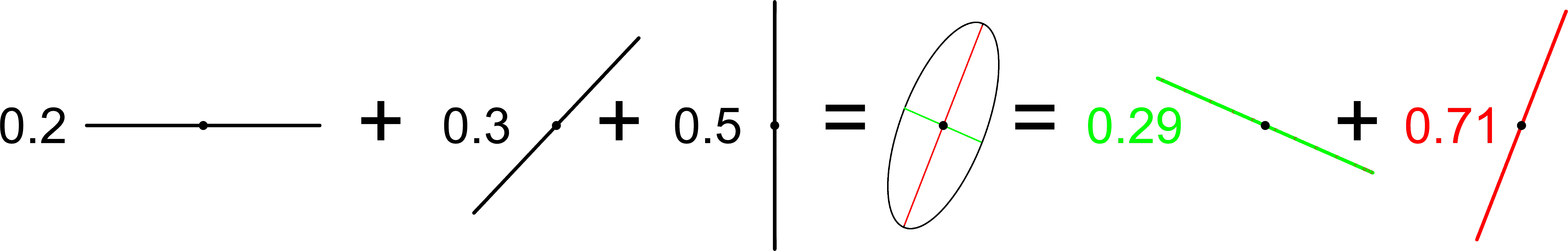}
\caption{Two different dyad mixtures that lead to the same density matrix:
\protect\usebox{6}
\protect\usebox{7}. 
Matrices are depicted as ellipses and dyads are degenerate single axis ellipses.}
\label{f:mix}
\end{figure*}

\begin{figure}[t]
\begin{tabular}{cc}
\begin{minipage}{0.5\textwidth}
\begin{center}
\includegraphics[width=0.7\textwidth]{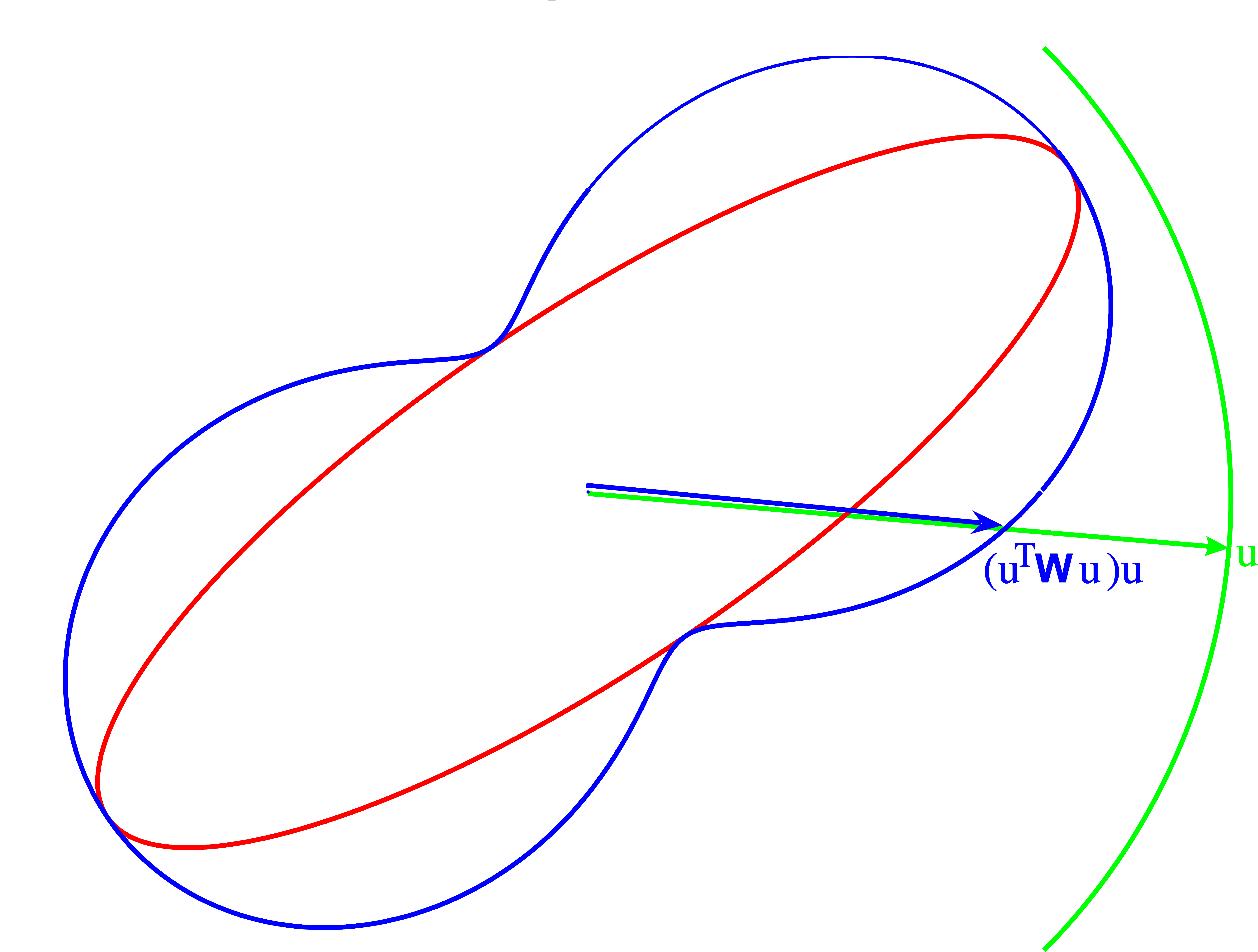}
\end{center}
\end{minipage}
&
\begin{minipage}{0.5\textwidth}
\begin{center}
\includegraphics[width=0.7\textwidth]{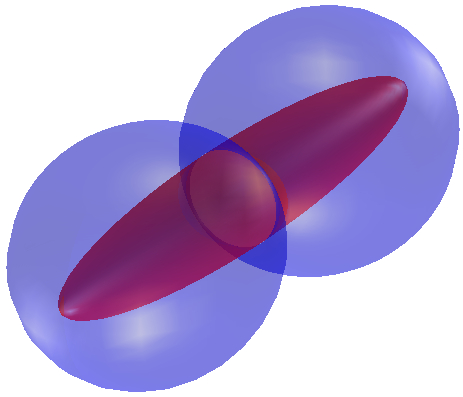}
\end{center}
\end{minipage}\\
(a) & (b)
\end{tabular}
\begin{minipage}{1\textwidth}
\caption{Figure (a) depicts a red ellipse $\{\W\u: \lVert\u\rVert_2=1\}$ for some density matrix
$\W$. The green curve shows part of the unit ball. The blue figure-eight
is a plot of the
generalized probabilities in direction $\u$, i.e. 
$\tr(\W\u\u^{\top}) \u$. 
Figure (b) plots a 3-dimensional density matrix (red ellipsoid) and its associated 
generalized probability surface (in blue).}
\label{f:fig8}
\end{minipage}
\end{figure}

Density matrices generalize finite probability distributions.
They can be defined as mixtures of dyads 
$\W=\sum_i \omega_i \w_i \w_i^\top$ 
where the mixture coefficients $\omega_i$ are non-negative and 
sum to one.
There may be an arbitrary number of components in the
mixture. However, any $n$ dimensional density matrix
can be decomposed into a mixture of $n$ orthogonal {\em eigendyads},
one for each eigenvector (see Figure \ref{f:mix}).
Mixtures of dyads are always symmetric%
\footnote{In quantum physics complex numbers are used instead of
reals. In that case ``symmetric'' is replaced by
``hermitian'' and all our formulas hold for that case as well.} 
and positive definite. A density matrix $\W$
can be depicted as an ellipse which is an
affine transformation of the unit ball: $\{\W\u: ||\u||_2=1\}$ 
(See Figure \ref{f:fig8}).
A dyad is a degenerate ellipse with a single axis in direction
$\pm\u$ that has radius one (Figure \ref{f:mix}). 
Note that dyads have trace one:
$$\tr(\u\u^{\top})=\tr(\u^{\top}
\u)=\lVert \u \rVert^2_2=1.$$ 
Therefore, density matrices also have trace one.

A density matrix $\W$ assigns generalized probability
$\tr(\W \u\u^\top)$ to each unit vector $\u$ 
and its associated dyad $\u\u^\top$ (see Figure
\ref{f:fig8}).
This probability is independent of how $\W$ is
expressed as a mixture and
can be rewritten as
$\u^\top\W\u$. 
Note that if the symmetric positive definite matrix 
$\A$ is viewed as a covariance matrix of a random cost vector $\c$, 
then $\u^\top\A\u$ is the variance of the cost along direction $\u$,
i.e. the variance of $\c \cdot \u$.

\sbox{6}
{
${\scriptstyle 0.2}\left(\begin{smallmatrix}1\\0\end{smallmatrix}\right)(\begin{smallmatrix}1&0\end{smallmatrix})+ 
{\scriptstyle 0.3}\left(\begin{smallmatrix}\sqrt{2}/2\\\sqrt{2}/2\end{smallmatrix}\right)
(\begin{smallmatrix}\frac{\sqrt{2}}{2}&\frac{\sqrt{2}}{2}\end{smallmatrix}) + 
{\scriptstyle 0.5}\left(\begin{smallmatrix}0\\1\end{smallmatrix}\right)(\begin{smallmatrix}0&1\end{smallmatrix})=
\left(\begin{smallmatrix}
0.35& 0.15\\
0.15& 0.65
\end{smallmatrix}\right)=
$
}

\sbox{7}{
$\textcolor{DarkGreen}{{\scriptstyle 0.29}\left(\begin{smallmatrix}-0.92\\\;\;0.38\end{smallmatrix}\right)
(\begin{smallmatrix}-0.92&0.38\end{smallmatrix})}+
\textcolor{red}{{\scriptstyle 0.71}\left(\begin{smallmatrix}0.38\\0.92\end{smallmatrix}\right)
(\begin{smallmatrix}0.38&0.92\end{smallmatrix})}$
}

If $\aa=(\alpha_1,\ldots,\alpha_n)$ is a probability vector,
then the $n$-dimensional matrix $\diag(\aa)$
with vector $\aa$ as its diagonal is a density matrix.
Note that $\diag(\aa) = \sum_i \alpha_i \e_i \e_i^\top$, 
where the $\e_i$ are the standard basis vectors.
Thus conventional probability distributions are special density matrices
where the eigensystem is restricted to be the identity matrix. 
In this paper we develop a Bayesian style analysis for the
case when the eigensystem is allowed to be arbitrary.

\begin{figure}[t]
\begin{tabular}{cc}
\begin{minipage}{0.5\textwidth}
\begin{center}
\includegraphics[width=0.5\textwidth]{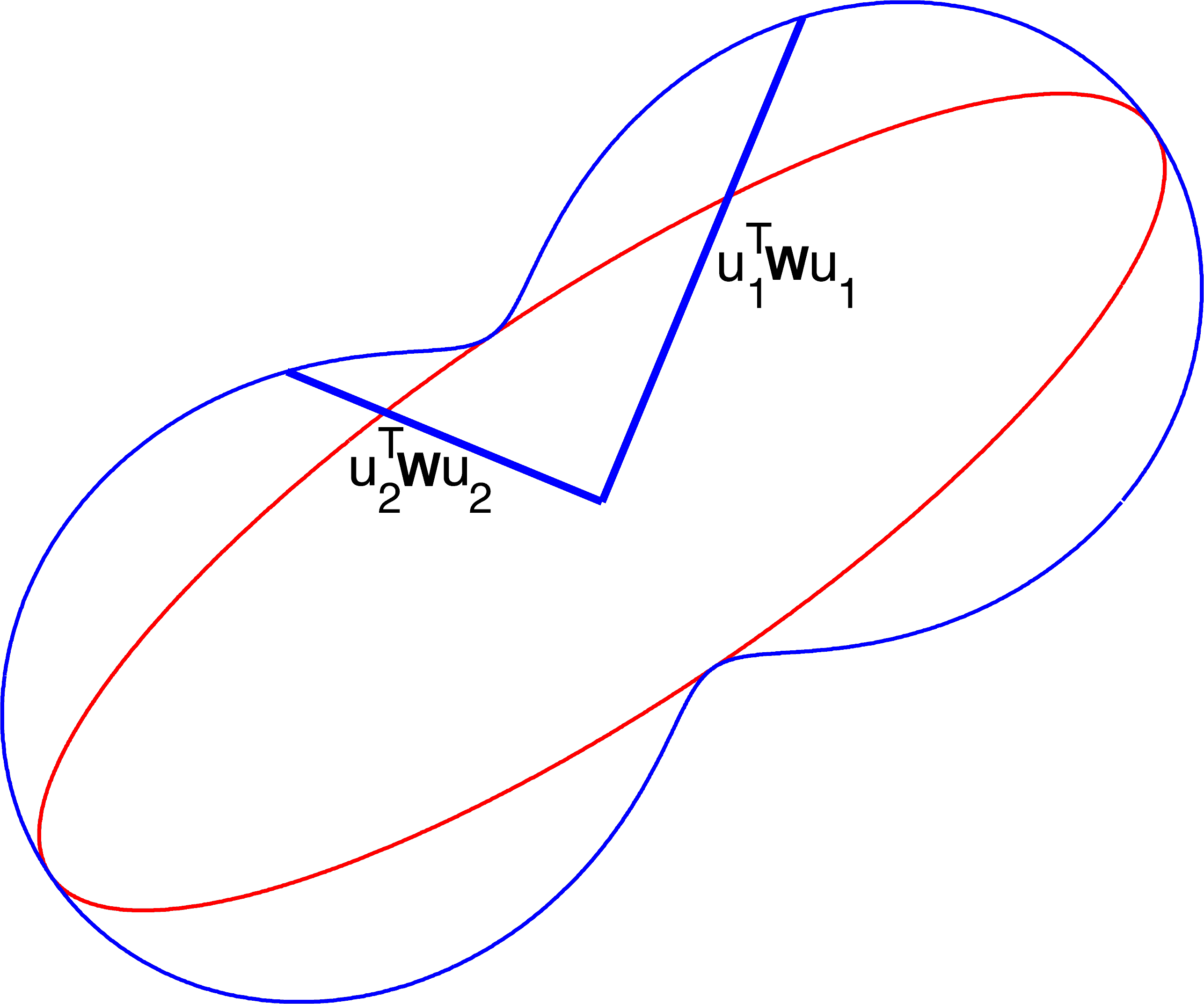}
\end{center}
\end{minipage}
&
\begin{minipage}{0.5\textwidth}
\begin{center}
\includegraphics[width=0.5\textwidth]{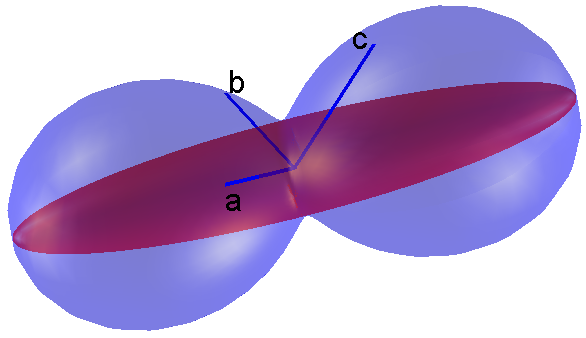}
\end{center}
\end{minipage}\\
(a) & (b)
\end{tabular}
\begin{minipage}{1\textwidth}
\caption{For a set of orthogonal directions $\u_i$ and a density matrix $\W$, the sum of generalized
probabilities $\tr(\W\u_i\u_i^\top)$ over the set is one. Figure (a) shows this for 2-dimensional
case: the red ellipse is a density matrix $\W$, 
the blue figure-eight is a plot of
the generalized probablity $\tr(\W \u\u^\top)$ around the circle, and
for any two orthogonal vectors $\u_1$ and $\u_2$, 
$\tr(\W\u_1\u_1^\top)+\tr(\W\u_2\u_2^\top) =1$.
Figure (b) shows the three-dimensional case: for any
three orthogonal directions $\u_1$, $\u_2$ and $\u_3$,
the probabilities $a$, $b$ and $c$ of the three associated 
dyads sum to one.}
\label{f:orth}
\end{minipage}
\end{figure}

Perhaps the simplest case to see that something unusual is
going on is the uniform density matrix, 
i.e. $\frac{1}{n}$ times identity $\I$. This
density matrix assigns probability $\frac{1}{n}$ to every unit vector, 
even though there are infinitely many of them. 
However, note that the sum of generalized probabilities of any
set of $n$ orthogonal dyads is $n \frac{1}{n}=1$.
As a matter of fact for any density matrix $\W$ and
any set of $n$ orthogonal directions $\u_i$,
the total generalized probability is one 
(see Figure \ref{f:orth})
\begin{equation}
\label{e:probsum}
\sum_{i=1}^n \tr(\W \;\u_i \u_i^\top)
=
\tr( \W \;
\underbrace{\sum_i \u_i\u_i^\top}_\I
)=\tr(\W)=1.
\end{equation}
This means that while in the conventional case probabilities are additive
over the points in the set, in the generalized case
probabilities are additive over orthogonal sets of dyads.

In this paper we use density matrices as generalized priors
and develop a unifying Bayesian probability calculus
for density matrices with rules for translating between
joints and conditionals.
All formulas retain the conventional case as the special case
when the matrices are diagonal. 
In previous work \cite{\qbayes} we derived a generalized Bayes rule
based on the minimum relative entropy principle,
but no satisfactory probabilistic interpretation was given
for this rule. This Bayes rule fits nicely into our new
calculus and we can interpret it using the
notion of generalized probability introduced above.

For any fixed orthonormal system $\u_i$,
one can use the dyads $\u_i\u_i^\top$ as elementary
events of a conventional probability space. As already discussed,
any density matrix can be seen as assigning conventional probabilities to these
events that sum to one. Thus if the orthonormal system is fixed, generalized probability space
is reduced to conventional probability space over the vectors in the chosen system.
Our approach is fundamentally different in that we 
use density matrices to maintain uncertainty
over all orthonormal systems. Our conditional density matrices
are part of the probabilistic system specified by a generalized 
joint probability distribution. In particular, our conditioning
method leads to generalizations of the theorem of total probability
that involve density matrices.

In \cite{\meg} various on-line learning updates were
generalized from vector parameters to matrix
parameters. Following \cite{\expgrad}, the
updates were derived by minimizing the loss on the current instance
plus a divergence to the last parameter.
In this paper we use the same method
for deriving a Bayes rule for density
matrices, which becomes the foundation of our generalized
probability calculus.
When the parameters are probability vectors
over the set of models, then the ``conventional'' Bayes
rule can be derived using the 
relative entropy as the divergence
(e.g. \cite{zellner98,\dotpred,\myspeech}).
Analogously, we now use the quantum
relative entropy, introduced by Umegaki,
to derive the generalized Bayes rule.

The new rule uses matrix logarithms and exponentials
to avoid the fact that symmetric positive definite matrices
are not closed under the matrix product.
The rule is strikingly similar to the conventional Bayes rule
and retains the latter
as a special case when the matrices are diagonal.
Various cancellations occur when the conventional Bayes rule 
is applied iteratively
and as we shall see, similar cancellations happen with the new rule
(See Section \ref{s:chain}). 
The conventional Bayes rule
may be seen as a soft maximum calculation
and the new rule as a soft calculation of the eigenvector 
with the largest eigenvalue (see figures \ref{f:bayes} and \ref{f:worm}).
In figures \ref{f:sigmadiag} and \ref{f:sigmarot} we plot
the projections of posterior onto the eigendirections
of the fixed datalikelihood matrix $\D(y|\MS)$. The
projection onto the eigendirection of the largest
eigenvalue is a sigmoid like function.

The mathematics applied in this paper
are most commonly used in quantum physics. 
For example, the assignment of generalized probabilities $\tr(\W\u\u^\top)$, can be
seen as the outcome of a quantum measurement of a system 
in mixture state $\W$ being acted upon by a measurement
apparatus described by the dyad $\u\u^\top$.
It is tempting to call the new rule the
``quantum Bayes rule''. 
However, we currently do not have a quantum physical interpretation of this rule. 
In particular, the state collapse following a measurement does not explicitly appear in our calculus, also
our Bayes rule can not be described as a unitary evolution of the prior state.
The term ``quantum Bayes rule'' also has been
claimed before in \cite{sbc-qbr-01}, where they derive a rule
that describes 
uncertainty information about unobserved quantum measurements
of a composite system as a density matrix.

Our work is most closely related to a paper by Cerf and Adami \cite{ca-qecp-99}, 
where, in the context of quantum information
theory, a formula was
proposed for the conditional density matrix that uses the
matrix exponential and matrix logarithm. 
This special formula appears in our calculus and is now 
put in a more general context.
We hope to transfer many techniques
developed in Bayesian Statistics based on the conventional Bayes rule 
to the context of generalized probabilities.

The paper is organized as follows. 
Section \ref{s:matrix} recalls the relevant matrix algebra facts.
Section \ref{s:gleason} introduces density matrices and generalized probability distributions and states
Gleason's theorem that establishes an equivalence between them. 
Then, in Section \ref{s:odot} we introduce a generalization $\odot$ of the matrix product that is commutative and preserves positive
definiteness. This $\odot$ operation is central to our calculus. 
Section \ref{s:joint}
introduces generalized joint distributions. Section \ref{s:marginal} discusses marginalizing the joints.
Next, in Section \ref{s:cond} we give formulas for conditional density matrices. Section 
\ref{s:total} presents generalizations of the Theorem of Total Probability. 
In Section \ref{s:bayes} we present the founding piece of this work, the
Bayes rule for density matrices, its derivation and various properties.
We also discuss how the new Bayes rule for density
matrices is in some sense the conventional Bayes rule in an
optimally chosen eigensystem. Section \ref{s:formulas} summarizes all the rules in our calculus and
their justifications. In the conclusion section 
we discuss again how our new calculus relates to quantum physics and possible generalizations of it.

\section{Facts on Matrices and Basic Notation}
\label{s:matrix}
In this paper generalized probability distributions, conditionals and data likelihoods are represented
as symmetric positive definite matrices. We will now discuss some relevant matrix algebra facts.

The basic fact that we use a lot is the eigendecomposition of symmetric matrices: 
\[\S = \SS\boldsymbol{\sigma}\SS^\top = \sum_{i=1}^n \sigma_i\,\s_i\s_i^\top \]
This says that every such matrix
can be written as a product of an orthogonal matrix of eigenvectors $\SS$ times 
a diagonal matrix of eigenvalues $\boldsymbol{\gamma}$ times $\SS^\top$.
Alternatively it can be written as mixture of eigendyads 
formed from the eigenvectors where the eigenvalues
act as mixture coefficients.

Any symmetric positive definite\footnote{%
We use the convention that positive definite matrices
have non-negative eigenvalues and {\em strictly} positive
definite matrices have positive eigenvalues.}
matrix $\C$ can be seen as a covariance matrix of some random cost
vector $\c \in \R^n$, i.e. $\C = \E\left((\c-\E(\c)(\c-\E(\c))^\top\right)$.
A covariance matrix $\C$
can be depicted as an ellipse
$\{\C\u:\lVert\u\rVert_2 = 1\}$ centered at the origin,
where the eigenvectors form the principal 
axes and the eigenvalues are the radii
of the axes (see Figure \ref{f:fig8}).

Note that a covariance matrix $\C$ is
diagonal if the components of the cost vector are independent.
The variance of the cost vector $\c$ along a vector $\u$, 
that is the variance of the dot
product $\c^\top\u$, has the form
\begin{equation*}
\begin{split}
\V(\c^{\top}\u) =& \E\left((\c^{\top}\u - \E(\c^{\top}\u))^2\right)\\
=&\E\left(((\c^{\top} - \E(\c^{\top}))\u)^{\top}((\c^{\top} - \E(\c^{\top}))\u)\right)\\
=&\E\left(\u^{\top}(\c-\E(\c))(\c - \E(\c))^{\top})\u \right)\\
=& \u^{\top}\C\u.
\end{split}
\end{equation*}
The variance along an eigenvector of the covariance matrix is the
corresponding eigenvalue.
Using this interpretation, the 
matrix $\C$ may be seen as a mapping
from the unit ball to $\R_{\geq 0}$, 
i.e. unit vector $\u$ is mapped to $\u^\top\C \u$. 
Figure \ref{f:fig8} depicts the resulting figure-8-like plots in 2 
and 3 dimensions. A second interpretation of the scalar
$\u^\top\C\u$ is the square length of $\u$
w.r.t. the basis $\sqrt{\C}$, that is
$\u^\top\C\u= \u^\top\sqrt{\C}\sqrt{\C}\u=
\lVert\sqrt{\C}\u\rVert_2^2$.

The trace $\tr(\E)$ of an arbitrary square matrix $\E$
is the sum of its diagonal elements $\E_{ii}$. It is a linear operator.
Recall that $\tr(\E\F)=\tr(\F\E)$ for
any matrices $\E\in\R^{n\times m},\;
\F\in\R^{m\times n}$. Also, for symmetric square matrices, $\tr(\S\T) = \sum_{i,j} S_{ij}T_{ij}$,
thus trace can be seen as a dot product between matrices. 
The trace has a useful cycling property: 
for arbitrary matrices $\E,\F,\G$ with compatible dimensions $\tr(\E\F\G) = \tr(\F\G\E) = \tr(\G\E\F)$.
From this follows that trace is {\em rotation invariant}
in the sense that for any orthogonal matrix $\UU$,
$\tr(\UU\E\UU^\top) =\tr(\UU^\top\UU\E)=\tr(\E).$
If $\S$ is symmetric, setting $\UU$ to be the eigensystem
of $\S$ results in the observation that trace is equal 
to the sum of eigenvalues of a matrix.
Also, for any orthogonal system\footnote{%
A set of unit vectors $\u_i$ is orthogonal iff
$\sum_i \u_i\u_i^\top = \I$.}
$\u_i$, 
$$\tr(\S)=\tr(\underbrace{\sum_{i=1}^n\u_i\u_i^\top}_{\I}\S)
=\sum_{i=1}^n \u_i^\top\S\u_i.$$
Therefore if $\S$ is symmetric positive definite,
then $\tr(\S)$ is the total variance
along any set of orthogonal directions.
Recall that density matrices have trace one 
and therefore in this case this total variance is always one
(See Figure \ref{f:orth}).

The matrix exponential $\expm(\S)$ 
of the symmetric matrix $\S=\sum_i \sigma_i\,\s_i\s_i^\top$
is computed by exponentiating the eigenvalues and leaving the eigenvectors unchanged:
$\expm(\S) = \sum_i \exp(\sigma_i)\,\s_i\s_i^\top$. 
The matrix logarithm $\logm(\A)$ is
defined similarly but now $\A$ must
be strictly positive definite. Clearly, the two functions
are inverses of each other.
It is important to remember that 
$\expm\left(\S+\T\right) =\expm(\S)\expm(\T)$
only holds if $\S$ and $\T$ commute
i.e. $\S\T=\T\S$.%
\footnote{%
This occurs iff the two symmetric matrices
have the same eigensystem.}
However, the following trace inequality,
known as the Golden-Thompson
inequality%
\footnote{%
Note that the Golden-Thompson inequality does not
generalize to three matrices, i.e. there exist symmetric
$\S$, $\T$, $\U$, s.t.
$\tr(\expm(\S)\expm(\T)\expm(\U)) \ngeq 
\tr(\expm\left(\S+\T+\U\right))$.}
\cite{\bhatia}, always holds:
\begin{equation}
\label{e:gt}
\tr(\expm(\S)\expm(\T)) \geq \tr(\expm\left(\S+\T\right))
\text{ for symmetric $\S$ and $\T$,}
\end{equation}
where equality holds iff both symmetric matrices commute.

\section{Generalized Probability Distributions and Density Matrices}
\label{s:gleason}
In quantum physics a dyad $\u\u^\top$ represents a pure
state and density matrices are mixture states.
As we shall see density matrices can be interpreted as 
generalized probability distributions over the set of dyads.
Note that in this paper we want to address the statistics
community and use linear algebra notation
instead of Dirac notation.
Any probability vector $(P(M_i))$
can be represented as a diagonal matrix
$\diag(P(M_i))
=\sum_i P(M_i)\:\e_i\e_i^\top$,
where $\e_i$ denotes the $i$th standard basis vector.
This means that conventional 
probability vectors are special density matrices where
the eigenvectors are fixed to be the standard basis vectors.

For the sake of simplicity we assume that
our vector space is $\R^n$. However,
everything discussed in this section holds 
for separable finite or infinite dimensional real and
complex Hilbert spaces.

A function $\mu(\u)$ from unit vectors $\u$ in $\R^n$ to $\R$ is called a \textit{generalized 
probability distributions} if the following two conditions hold:
\begin{itemize}
\item $\forall \u$, $0 \leq \mu(\u) \leq 1$.
\item If $\u_1, \dotsc, \u_n$ form an orthonormal system for
$\R^n$, then $\sum \mu(\u_i) = 1$.
\end{itemize}

Gleason's Theorem states that there is a one-to-one correspondence 
between generalized probability distributions
and density matrices\footnote{%
The core of the original proof of Gleason's Theorem was for $\R^3$
\cite{gle},
and he then generalized the proof to separable
real and complex Hilbert spaces of dimension $n\geq 3$.}
in $\R^{n \times n}$:
\begin{theorem}\cite{gle}
Let $n \geq 3$.\footnote{%
A slightly different version of this theorem that is
based on ``effects'' instead of dyads holds for dimension
2 as well \citep{fuchs-effects}.}
Then any generalized probability distribution $\mu$ on $\R^n$ has the form
$\mu(\u) = \tr(\W\u\u^{\top})$,
for a uniquely defined density matrix $\W$. 
\end{theorem}
It is easy to see that every density matrix defines a
generalized probability distribution. 
The other direction, is highly non-trivial.\footnote{%
However, if dyads are replaced by ``effects'' 
then the proofs are much simpler \citep{fuchs-effects}.}
As discussed in the introduction, the dyads $\u\u^\top$
function as elementary events. One may ask what corresponds to
arbitrary events and how probabilities are defined for
them. In the conventional case, an event is a subset of the
domain which can be represented as a vector in $\{0,1\}^n$.
In the generalized setting, an event is a symmetric positive
definite matrix $\P$ with eigenvalues in $\{0,1\}$. Each
such matrix $\P$ with eigendecomposition $\sum_{i=1}^k \p_i\p_i^\top$ 
is a projection matrix for a subspace of
$\R^n$ and its probability w.r.t. a distribution $\W$ is
defined as the sum of the probabilities
of the elementary events $\p_i\p_i^\top$ comprising $\P$:
$$\tr(\W\P) 
=\sum_{i=1}^k \tr(\W \: \p_i\p_i^\top).$$
Interpreting $\P$ as a covariance matrix of some random variable, we can also expand 
$\W$ and sum the
variance along its eigendirections $\w_i$ weighted by
the eigenvalues $\omega_i$ which are probabilities:
\begin{equation}
\label{e:WP}
\tr(\W\P) 
= \tr(\sum_{i=1}^n \omega_i \;\w_i \w_i^\top \P)
= \sum_{i=1}^n
\underbrace{
   \overbrace{\omega_i}^{\text{probability}} 
   \overbrace{\w_i^\top \P \w_i}^{\text{variance}}
           }_{\text{expected variance}}
.\end{equation}

Random variables are defined in an analogous way.
In the conventional case a random variable associates a real
value with each point. 
Now a random variable is an arbitrary symmetric matrix $\S$.
Such matrices have arbitrary real numbers as their
eigenvalues and trace $\tr(\W\S)$ when $\S$ is expanded
becomes the expectation of the random variable
w.r.t. density $\W$:
\begin{equation}
\label{e:WS}
\tr({\W}{\S})
= \tr({\W}{\sum_i \sigma_i \s_i\s_i^\top})
=
\underbrace{\sum_i\overbrace{{\sigma_i}}^{\text{outcome}} 
\overbrace{{\s_i^\top}{\W}\:{\s_i}}^{\text{probability}}}_{\text{expected
outcome}}.
\end{equation}
As discussed before, the conventional case
of the expectation calculation
is always retained as a special case when all the
matrices are diagonal (i.e. fixed eigensystem $\I$).
In quantum physics the
expectation calculation $\tr(\W\S)$ has the following interpretation:
an instrument is represented by 
a hermitian matrix $\S$ and $\tr(\W\S)$
is the expected value of a {\em quantum measurement} of
the mixed state $\W$ with instrument $\S$.
The eigenvalues $\sigma_i$ of the instrument 
represent the possible numerical measurement outcomes.
Each one of those outcomes is observed with probability
$\s_i^\top \W \s_i$, where $\s_i$ is the
associated eigenvector of the instrument matrix $\S$.

In real quantum systems the measurement causes 
the mixtures state $\W$ to \textit{collapse} 
into one of the orthogonal states
$\{\s_1\s_1^\top,\ldots,\s_n,\s_n^\top\}$:
the successor state is $\s_i\s_i^\top$ with
probability $\s_i^\top\W\s_i$:
\[\W \;\;
\begin{array}{c} \text{\tiny  measurement}\\
                  \longrightarrow \\
                  \text{\tiny collapse}
\end{array}
\underbrace{\sum_i \overbrace{{\s_i^\top} {\W}\:
{\s_i}}^{\text{probability}}\;{\s_i\s_i^\top}}_{\text{expected
state}}.\]
As we shall see, the expected measurement calculations play an important part in
our calculus. However our update rules for density matrices
(such as our Bayes rule) do not explicitly include a collapse in the above sense.

Note that some of the equations above hold for arbitrary decompositions into
a linear combination of dyads of any size. 
For example \eqref{e:WP}, holds for any decomposition
$\W=\sum_i\omega_i\,\w_i\w_i^\top$, i.e. the $\omega_i$
may be negative, the $\w_i$ may
be non-orthogonal, and the size of the decomposition may be
larger than $n$.
If the $\omega_i$ are non-negative, then they form a
probability vector.
Similarly, \eqref{e:WS} also holds for any decomposition
$\S=\sum_i\sigma_i\s_i\s_i^\top$. However, quantum
measurements are always based on an orthogonal system.
Furthermore, orthogonal systems are special in that the orthogonal
decomposition of a density matrix $\W=\sum_i\omega_i\,\w_i\w_i^\top$ 
attains the minimum of the entropy $\sum_i - \omega_i \ln \omega_i$ over all
possible decompositions of $\W$ (Inequality (11.86) in \cite{NieChu00}).

A question that naturally arises is whether we can model
the generalized probability distributions defined above with
a conventional probability space. 
In other words,
is there a conventional probability space and two mappings:
one that maps density matrices to conventional probability distributions and the other mapping dyads to
events of this probability space. 
The requirement on these two mappings is that the conventional probability
calculations using the 
images of density matrices and dyads under these mappings satisfy the definition of the generalized 
probability distributions given above. 
Essentially, it is known that conventional probability spaces 
cannot satisfactorily model
generalized probabilities, but the details are rather
involved. This topic has received considerable attention in the quantum 
physics community and we refer readers to \cite{holevo} 
for an extended discussion of impossibility results. 
Here we only give one simple attempt
to model density matrices with a conventional probability
space and show that the two natural mappings 
fail to satisfy the requirements. 

A natural interpretation of a density matrix
is to view it as a parameterized density
over the unit sphere. We claim that if $\mu(\u)$ is the uniform
density on the sphere, then for any symmetric positive definite
matrix $\A\in \R^{n\times n}$ of trace $n$,
$\u^\top\A\:\u \:\mu(\u)$ is also a conventional probability density 
on the sphere:
\begin{eqnarray*}
\!\!\!\!\int \u^\top\overbrace{\sum_i \alpha_i \a_i
\a_i^\top}^{\A} \u \,\mu(\u) d \u
= \sum_i \alpha_i \int (\u^\top \a_i)^2 \mu(\u) d \u 
\\
= \tr(\A)\!\!
\int
(\u^\top\!(\frac{1}{\sqrt{n}},\ldots,\frac{1}{\sqrt{n}})^\top)^2 \mu(\u) d \u 
= \frac{\tr(\A)}{n}  \!\!
\underbrace{\int \overbrace{(\u)^2}^{1} \mu(\u) d \u}_1.
\end{eqnarray*}
In the second equality we used the fact that $\mu(\u)d \u$
is uniform and therefore the integral of 
$(\u^\top \a_i)^2$ is the same as the integral
of the squared dot product of $\u$ with uniform vector 
$(\frac{1}{\sqrt{n}},\ldots,\frac{1}{\sqrt{n}})^\top$.

We modeled density matrices as conventional probability
densities over the sphere. Now the natural mapping
from dyads to events in the conventional probability
space (the sphere) maps $\u\u^\top$ to $\{\u,-\u\}$.
However the probability of the latter sets of size 2
is zero with respect to the conventional probabilities densities
we defined on the sphere. 
In particular the probability on any $n$ orthogonal
dyads does not sum to one.

\section{Commutative Matrix Product Operation}
\label{s:odot}

\begin{figure}[t]
\begin{tabular}{cc}
\begin{minipage}{0.5\textwidth}
\begin{center}
\includegraphics[width=1\textwidth]{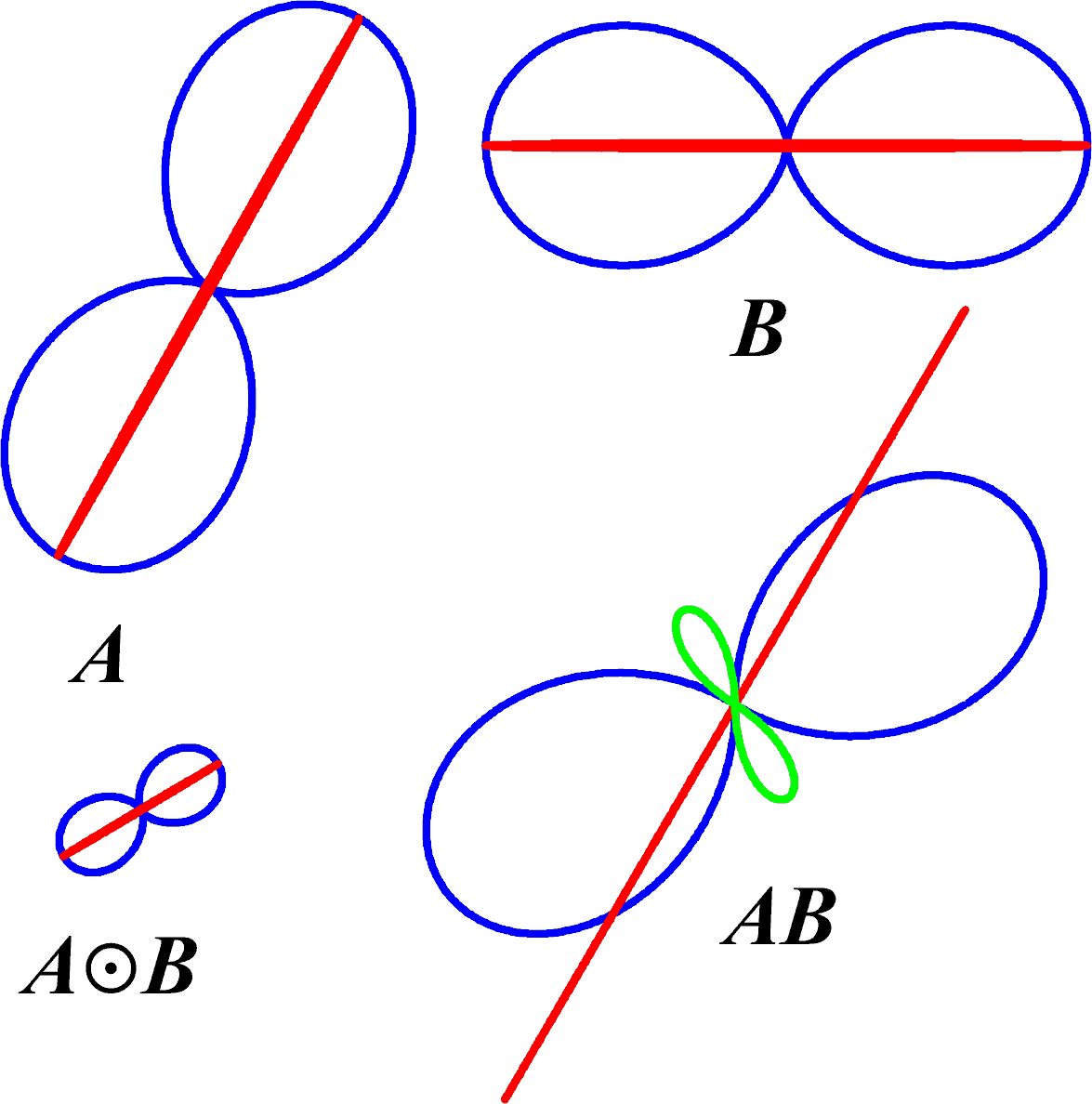}
\end{center}
\end{minipage}&
\begin{minipage}{0.5\textwidth}
\centerline{\includegraphics[width=0.6\textwidth]{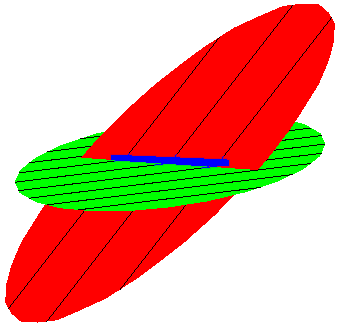}}
\end{minipage}\\
\begin{minipage}[t]{0.5\textwidth}
\caption{The matrix product of two positive 
definite matrices does not preserve positive definiteness. 
For two matrices $\A$ and $\B$ we plot their ellipses $\A\u,\B\u$ and figure eights 
$\tr(\A\u\u^{\top})\,\u, \tr(\B\u\u^{\top})\,\u$ (for unit $\u$). 
Both ellipses are very thin, i.e. the ratio between
the two eigenvalues of each matrix is 100.
We also plot the ellipse $\A\B\u$ 
and the curve $\tr(\A\B\u\u^{\top})\,\u$.
The latter curve consists of two figure eights, the larger one
constitutes the part where the trace is positive
and the smaller and skinnier one is the part where
the trace is negative.
This means that $\A\B$ is not positive definite any more.
The product is also not symmetric because the
min/max value of $\tr(\A\B\u\u^{\top})$
does not correspond to the axes of the ellipse.
Finally, the corresponding plots for $\A\odot\B$ 
indicate that this matrix is symmetric and positive definite.}
\label{f:odot}
\end{minipage}&
\begin{minipage}[t]{0.5\textwidth}
\caption{When the ellipses $\A$ and $\B$
don't have the same span, then $\A \odot \B$
lies in the intersection of both spans. In the depicted case the intersection
is a degenerate ellipse of dimension one
(blue line). This generalizes the following
intersection property of the matrix product 
when $\A$ and $\B$ are both diagonal (here of
dimension four):
$(\A\B)_{i,i} \ne 0 
\;\text{iff}\; \A_{i,i} \ne 0 \;\text{and}\; \B_{i,i} \ne 0 .
$
\protect\usebox{1}
}
\label{f:inters}
\end{minipage}
\end{tabular}
\end{figure}

It is well known that the product of two symmetric 
positive definite matrices
might be neither symmetric nor positive definite (see
Figure \ref{f:odot}). In this section we define
a commutative ``product'' operation between symmetric positive definite
matrices that does result in a symmetric positive definite matrix.
Our first definition of this operation requires the
two matrices to be strictly positive definite.
We then extend the definition to arbitrary symmetric positive definite matrices
and prove many properties of this product.

For two symmetric and strictly
positive definite matrices $\A$ and $\B$, we first define 
the $\odot$ as:
\begin{equation}
\label{e:explog}
\A \odot \B \;:=\; \overbrace{\expm(\overbrace{\logm \overbrace{\A}^{\text{sym.pos.def.}}}^{\text{sym.}} + 
\overbrace{\logm \overbrace{\B}^{\text{sym.pos.def.}}}^{\text{sym.}})}^{\text{sym.pos.def.}}, 
\end{equation}
where here the exponential and logarithm
are matrix functions.
The matrix log of both matrices produces
symmetric matrices which are closed under addition
and the matrix exponential
of the sum returns a symmetric positive definite matrix. 
See Figure \ref{f:odot} for a comparison of matrix product and $\odot$.

Note that we expressed the operation $\odot$ between symmetric strictly positive
definite matrices as a $+$ operation between
symmetric matrices. Similarly, for any two arbitrary
symmetric matrices $\S$ and $\T$,
$$\S+\T =\logm(\expm(\S)\odot\expm(\T)).$$

The operation $\odot$ was used in \cite{alexa}
to define a ``product'' between two linear transformations
that is commutative.
In this paper we use $\odot$ to define 
conditional density matrices and generalizations of the Bayes rule. 
A similar path was followed by \cite{ca-qecp-99}
for defining conditional density matrices of composite systems.
We also give a motivation for the operation based on
the minimum relative entropy principle (as was done in the conference
paper \citep{\qbayes}) and our probability calculus
includes the formula of \citep{ca-qecp-99} for composite
systems as a special case.

\sbox{1}
{
\begin{minipage}[t]{0.5\textwidth}
\begin{tabular}{cc|c}
$\diag(\A)$&$\diag(\B)$&$\diag(\A\B)$\\
\hline
0&0&0\\
a&0&0\\
0&b&0\\
a&b&ab
\end{tabular}
\end{minipage}
}

\begin{figure*}[t]
\begin{center}
\includegraphics[width=1\textwidth]{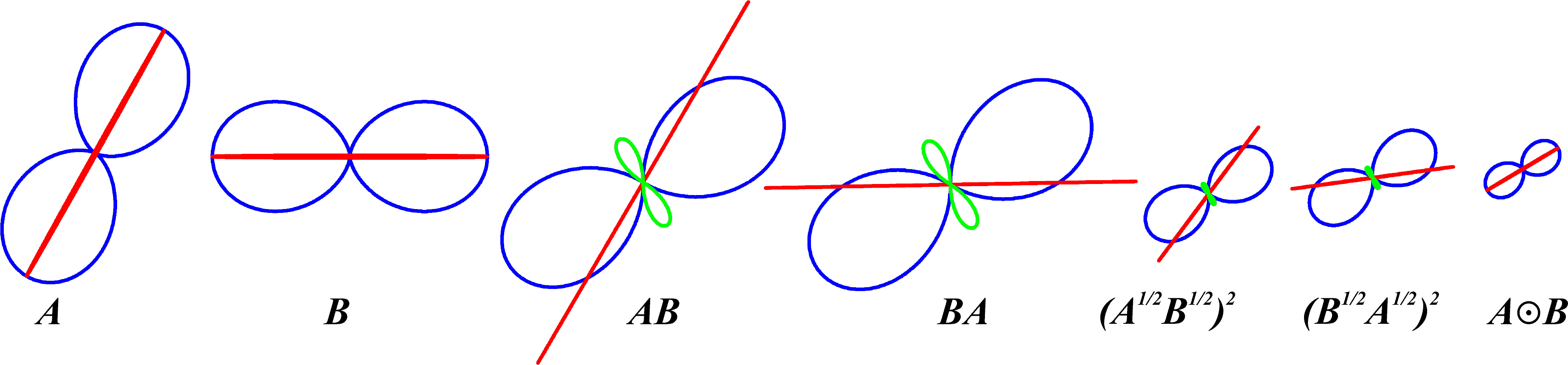}
\end{center}
\caption{The behavior of the limit formula for $\odot$
operation. We can see that the additional figure eights
indicating negative definiteness are smaller for
$(\A^{1/2}\B^{1/2})^2$ than for $\A\B$.
As $n$ increases, the additional figure eights 
shrink further and $\lim_{n\rightarrow\infty} 
(\A^{1/n}\B^{1/n})^n= \A\odot\B$ becomes
positive definite. Also, $\A\B$ and $\B\A$
are fairly different from one another. The matrices $(\A^{1/2}\B^{1/2})^2$ and $(\B^{1/2}\A^{1/2})^2$
are already more similar and the difference between the two multiplication orders decreases
with $n$ until in the limit $\A\odot\B = \B\odot\A$.}
\label{f:lim}
\end{figure*}

Note that the formula for $\odot$ in Equation \eqref{e:explog} is not defined if some of
the eigenvalues of $\A$ or $\B$ are zero. We now rewrite the operation
using the Lie-Trotter formula and then extend it
to arbitrary positive definite matrices.
The Lie-Trotter formula (see e.g. \cite{\bhatia}) 
is the following equation: 
$$\expm(\E+\F)=
\lim_{n\rightarrow \infty}
\left(\expm(\E/n) \expm(\F/n)\right)^n,\quad
 \text{any square matrices $\E,\F$.}
$$
By choosing $\E=\logm \A$ and $\F=\logm \B$,
for symmetric and strictly positive definite $\A$ and
$\B$, we obtain:
\begin{equation*}
\expm(\logm \A + \logm \B) =\lim_{n\rightarrow\infty} 
(\A^{1/n}\B^{1/n})^n.
\end{equation*}
As $n$ increases, $(\A^{1/n}\B^{1/n})^n$
gets closer and closer to being positive definite
and symmetric. 
The first couple iterations of the limit formula are plotted in Figure \ref{f:lim}. 
See \cite{alexa} for additional plots.
Notice that the limit is defined even when $\A$ and $\B$
have zero eigenvalues.
We therefore extend the definition of $\odot$ to arbitrary
symmetric positive definite matrices $\A$ and $\B$:
\begin{equation}
\label{e:odotlim}
\A \odot \B
:=\lim_{n\rightarrow\infty} (\A^{1/n}\B^{1/n})^n.
\end{equation}

From now on we use the above exended definition of $\odot$. 
Numerous properties of this operation are given below.
\begin{theorem}
\label{t:items}
For any symmetric positive definite matrices $\A,\B,\C$ the following holds:
\begin{enumerate}[label=OP\arabic*.,ref=OP\arabic*]
\item \label{i:inters}Intersection property:
$$\range(\A \odot \B) \;=\; \range(\A)\cap\range(\B).$$
where the range of a matrix is the linear subspace spanned
by the columns of the matrix.
This property generalizes the intersection properties
for products of diagonal matrices
(which model conventional probability distributions):
the product of two diagonal matrices
with the characteristic vectors of two subsets as diagonals
gives a diagonal matrix formed from
the characteristic vector of the intersection
(See Figure \ref{f:inters}).
\item \label{i:logplus}
Let $\RR_{\A}$ be a matrix whose columns form an orthonormal basis for the range of $\A$, i.e.
$\RR_{\A} \in \R^{n \times k}$ and
$\RR_{\A}^{\top}\RR_{\A} = \I_k$, where $k$ is 
the dimensionality of the
range of $\A$.
Define $\RR_{\B}$ analogously. In a similar fashion
$\RR_{\A\cap\B}$ will contain the basis for the intersection of ranges. 
Let $\logm^+$ denote the modified matrix logarithm that takes the log of non-zero
eigenvalues but leaves the zero eigenvalues unchanged. This operation can be also defined by the
following formula:%
\footnote{%
Note that when the rank $k$ of $\A$ is zero, then
one still can define the projections in a consistent
manner. In this case $\RR_\A$ is of dimension $n\times 0$,
and the matrices $\RR_\A^\top\RR_\A$ and $\logm(\RR_\A^\top\E \RR_\A)$  
are of dimension $0 \times 0$ for any $\E \in \R^{n\times
n}$. Also it is natural to define $\RR_\A\E\RR_\A^\top$
as the $n\times n$ zero matrix $\zero$.
With this definition, the r.h.s. of \eqref{e:logplus}
is $\zero$ when $\A$ is $\zero$.}
\begin{equation}
\label{e:logplus}
\logm^+\A =
\RR_{\A}\;\logm(\RR_{\A}^{\top}\A\RR_{\A})\;\RR_{\A}^{\top}.
\end{equation}
With this notation, $\odot$ can be written as
\begin{equation}
\label{e:odotfancy}
\A \odot \B =
\RR_{\A\cap\B} \;\expm(\RR_{\A\cap\B}^{\top}
(\logm^+\!\!\A+\logm^+\!\B)\RR_{\A\cap\B})\;\RR_{\A\cap\B}^{\top}.
\end{equation}

\item \label{i:prodgen}$\A\odot\B = \A\B$ if $\A$ and $\B$ commute.
\item $\odot$ is commutative, i.e. $\A \odot \B = \B \odot \A$.
\item The identity matrix is the neutral element, i.e.
$\A\odot\I = \A$.
\item $(c\A)\odot\B = c(\A\odot\B)$, for any scalar $c>0$.
\item $\A\odot\A^{-1} = \I$ for invertible
$\A$. Also, $\A\odot\A^{+} = \P_{\A}$, where $\A^{+}$ denotes the 
pseudoinverse and $\P_{\A}$ is the projection
matrix\footnote{%
Note that $\P_{\A}=\RR_{\A}\RR_{\A}^\top$.}
for $\range(\A)$.
\item\label{i:assoc} 
$\odot$ is associative, i.e. $(\A \odot \B) \odot \C = \A \odot (\B \odot \C)$.
\item\label{i:monot} 
Monotonic convergence of the limit defining $\odot$:
\[\forall n \geq 1: \;\tr(\A^{1/(n+1)}\B^{1/(n+1)})^{n+1} \leq 
\tr(\A^{1/n}\B^{1/n})^n\].

\item\label{i:ubound} $\tr(\A\odot\B) \leq \tr(\A\B)$,
where equality holds iff $\A$ and $\B$ commute. In particular, for any unit $\u$ 
$\tr(\A\odot\u\u^\top)=\tr(\A\u\u^\top)$ iff $\u$ is an eigenvector of $\A$.

\item\label{i:rpinch} 
For any unit direction $\u\in \range(\A)$,
$\A\odot\u\u^{\top} = e^{\u^{\top}(\logm^+\A)\u}\,
\u\u^{\top}$. 
\item\label{i:average}
For any unit direction $\u$ 
and eigendecomposition $\sum_i
\alpha_i\a_i\a_i^\top$ of a strictly positive definite
matrix $\A$,
$$\tr(\A\u\u^\top) = \sum_i (\u^\top\a_i)^2 \alpha_i,
\text{  and  }
\tr(\A\odot\u\u^\top)  = \prod_i \alpha_i^{(\u^\top\a_i)^2},
$$
i.e. the matrix product corresponds to an arithmetic
average and the $\odot$ product to a geometric average
of the eigenvalues of $\A$.
\item\label{i:det} $\det(\A\odot\B) = \det(\A)\det(\B)$, 
which is the same as for the normal matrix product.

\item\label{i:pinchdet} For any orthogonal system $\u_i$, we have 
$\prod_i \tr(\A\odot\u_i\u_i^\top) = \det(\A)$.

\item\label{i:prodpinch} 
For any unit direction $\u$,
$\tr((\A\odot\B)\odot\u\u^\top) =
\tr(\A\odot\u\u^\top)\;\tr(\B\odot\u\u^\top)$.

\item\label{i:pinchinverse} 
For any unit direction $\u\in \range{\A}$,
$\tr(\A^{+}\odot\u\u^\top) = \frac{1}{\tr(\A\odot\u\u^\top)}$,
where $\A^+$ denotes the pseudoinverse.
\end{enumerate}
\end{theorem}

\begin{proof}
Properties \ref{i:inters} and \ref{i:logplus} follow from results in \cite{kato78} 
or Theorem 1.2 of \cite{s-fiqp-79}. 
Here we only prove that $\range(\A\odot\B) \subseteq
\range(\A)\cap\range(\B)$.
We can split the limit defining $\odot$ as follows:
\begin{equation}
\label{e:split}
\A \odot \B = \lim_{n\rightarrow\infty} 
(\A^{1/n}\B^{1/n})^n = \lim_{n\rightarrow\infty}\A^{1/n} 
\lim_{n\rightarrow\infty} \B^{1/n}(\A^{1/n}\B^{1/n})^{n-1}.
\end{equation}
Here we used the property that 
$\lim \E_n\F_n = \lim \E_n \lim \F_n$ if all the limits exist. This follows
from the corresponding sum and product properties of scalar limits and the fact that entries of a product 
matrix are finite sums of products.

It is easy to see that $\lim_{n\rightarrow\infty}\A^{1/n} = \P_{\A}$
because the matrix power for a symmetric matrix 
corresponds to taking powers of the eigenvalues 
and $n$-th roots converge to either zero or one. 
Thus the limit is a matrix whose 
eigenvalues are $0$ or $1$, which is a projection matrix. 
By plugging 
$$\lim_{n\rightarrow\infty}\A^{1/n} = \P_{\A}=  
\P_{\A}\P_{\A}= \P_{\A}\lim_{n\rightarrow\infty}\A^{1/n}$$ 
into \eqref{e:split} we get
\[\A \odot \B = \P_{\A}
\lim_{n\rightarrow\infty}\A^{1/n} 
\lim_{n\rightarrow\infty} \B^{1/n}(\A^{1/n}\B^{1/n})^{n-1}
= \P_{\A} (\A \odot\B).\]
This implies that $\range(\A\odot\B) \subseteq \range(\A)$.
Similarly we can prove $\A \odot \B = (\A \odot \B) \P_{\B}$,
which implies that $\range(\A\odot\B) \subseteq \range(\B)$,
and therefore
$\range(\A\odot\B) \subseteq \range(\A)\cap\range(\B)$. 

Property \ref{i:prodgen} can be seen from the definition of $\odot$ via the limit formula \eqref{e:odotlim}: when $\A$ and $\B$ commute, 
then the $n$ copies of $\A^{1/n}$ in $(\A^{1/n}\B^{1/n})^n$ 
can be gathered into $\A$ and similarly for $\B^{1/n}$.

Properties 4 - 7 easily follow from the formula \eqref{e:odotfancy}
for $\odot$.

Property \ref{i:assoc}. 
For strictly positive definite $\A,\B,\C$ associativity reduces to the
associativity of addition in the log domain. 
To show it in general we use the representation \eqref{e:odotfancy}
of $\odot$ via the $\logm^+$ operation.
Let $\RR = \RR_{(\A\cap\B)\cap\C} = \RR_{\A\cap(\B\cap\C)}$. Then:
\begin{equation*}
(\A \odot \B) \odot \C =
\RR\;\expm(\RR^{\top}(\logm^+(\A\odot\B) + \logm^+\C)\RR)\;\RR^{\top}
\end{equation*}
Now we rewrite $\logm^+(\A\odot\B)$ using Equation \eqref{e:logplus}:
\begin{equation*}
\logm^+(\A\odot\B) 
= \RR_{\A\cap\B}\;\logm(\RR_{\A\cap\B}^{\top}(\A\odot\B)\RR_{\A\cap\B})
                \;\RR_{\A\cap\B}^{\top}
\end{equation*}
Substituting expression \eqref{e:odotfancy} 
for $\A\odot\B$ into the above 
and using $\RR_{\A\cap\B}^{\top}\RR_{\A\cap\B} = \I_k$ we get:
\begin{equation}
\label{e:plusAB}
\logm^+(\A\odot\B) =  
\underbrace{\RR_{\A\cap\B}\RR_{\A\cap\B}^{\top}}
_{\P_{\A\cap\B}}
(\logm^+\A+\logm^+\B)
\underbrace{\RR_{\A\cap\B} \RR_{\A\cap\B}^{\top}}
_{\P_{\A\cap\B}}
.
\end{equation}
Here $\P_{\A\cap\B}=\RR_{\A\cap\B}\RR_{\A\cap\B}^{\top}$ is
the projection matrix onto the subspace 
$\range(\A)\cap \range(\B)$. 
All the basis vectors of $\A\cap\B\cap\C$ obviously
lie in the larger subspace as well, thus the projection leaves them unchanged and we get 
$\P_{\A\cap\B} \RR = \RR$, $\RR^{\top}\P_{\A\cap\B}= \RR^{\top}$.
Thus:
\[(\A \odot \B) \odot \C =  
\RR\expm(\RR^{\top}(\logm^+\A +\logm^+\B + \logm^+\C)\RR)\RR^{\top}.\]
The same expression can be obtained 
for $\A\odot(\B\odot\C)$, thus establishing associativity.

Property \ref{i:monot}. By Fact 8.10.9 of \cite{matbook},
we have that for any positive definite matrices $\A$ and $\B$,
and $n\geq1$:
\[\tr(\A^n\B^n)^{n+1} \leq \tr(\A^{n+1}\B^{n+1})^n. \]
Now by substituting $\A=\A^{1/(n(n+1))}$, $\B=\B^{1/(n(n+1))}$ 
the monotonicity property \ref{i:monot} immediately
follows.

Property \ref{i:ubound}. 
When $\A$ and $\B$ are strictly positive definite, 
this inequality is an instantiation
of the Golden-Thompson inequality \eqref{e:gt}. 
For arbitrary positive definite matrices, the property follows from
the previous monotonicity property \ref{i:monot}. 
Note that there are symmetric positive definite
matrices $\A$, $\B$ and $\C$ s.t. 
$\tr(\A\odot\B\odot\C) \nleq \tr(\A\B\C)$.

Property \ref{i:rpinch}. We use the expression for $\odot$ operation given in Equation 
\eqref{e:odotfancy}. Since $\u\in\range(\A)$, the basis of the intersection space is $\u$ itself:
\[\u\u^{\top}\odot\A = \u\expm(\u^{\top}(\logm^+\u\u^{\top}
+ \logm^+\A)\u)\u^{\top}.\]
Note that $\logm^+\u\u^{\top} = \zero$ and that the
expression inside the exponential is a scalar.
The desired property immediately follows 
by moving this scalar to the front. 

Property \ref{i:average}.
The expression for the trace of the matrix product is the
expected measurement interpretation \eqref{e:WS} discussed in
Section \ref{s:gleason}. Note that $(\u^\top\a_i)^2$ is a probability vector and in this expression
$\u\u^\top$ can be replaced by any density matrix.

For the second trace $\tr(\A\odot\u\u^\top)$, we can rewrite it using \ref{i:rpinch} 
and eigendecomposition of $\A$ as follows:
\[\tr(\A\odot\u\u^\top) =  e^{\u^\top\logm\A\u} = e^{\sum_i (\a_i\cdot\u)^2\;\log\alpha_i}
= \prod_i \alpha_i^{(\u^\top\a_i)^2},\]
which is a weighted geometric average of $\alpha_i$ with
weights $(\u^\top\a_i)^2$.

Property \ref{i:det}. 
Since $\det(\E\F)=\det(\E)\det(\F)$, and for symmetric
matrices $\S$ and $r\in \R$, $\det(\S^r) = \det(\S)^r$, 
we have $\det((\A^{1/n}\B^{1/n})^n)=\det(\A)\det(\B)$
for all $n\in \N$.
By Property \eqref{e:odotlim},
the limit of the l.h.s. of the last equality becomes
$\A\odot\B$ and this proves the property.

Property \ref{i:pinchdet}. If $\A$ is not full rank, then $\det(\A)$ is zero. In that case, there
will be some $\u_i$ that is not in the range of $\A$. For that $\u_i$, $\tr(\A\odot\u_i\u_i^\top) = 0$, 
making the whole product zero.  When $\A$ has full rank, we rewrite the product as follows:
\begin{eqnarray*}
\prod_i \tr(\A\odot\u_i\u_i^\top) 
&\overset{\ref{i:rpinch}}{=}& \prod_i e^{\tr(\logm\A\u_i\u_i^\top})\\
&=& e^{\tr(\logm\A\sum \u_i\u_i^\top)} = e^{\tr(\logm\A)} = \prod_i \alpha_i = \det(\A).
\end{eqnarray*}

Property \ref{i:prodpinch}.
If $\u \notin \range(\A)\cap\range(\B)\overset{\eqref{i:inters}}{=}
\range(\A\odot\B)$, then the property trivially holds
because $\tr((\A\odot\B)\odot\u\u^\top)$ 
and either $\tr(\A\odot\u\u^\top)$ or $\tr(\B\odot\u\u^\top)$ are zero. 
When $\u \in \range(\A)\cap\range(\B)$, then
the property essentially follows from $e^{a+b}=e^a e^b$:
\begin{align*}
&\tr((\A\odot\B)\odot\u\u^\top) \\
\quad&\overset{\ref{i:rpinch}}{=} e^{\u^\top\logm^+(\A\odot\B)\u} 
\overset{\eqref{e:plusAB}}{=}
e^{\u^\top\P_{\A\cap\B}(\logm^+\A+\logm^+\B)\P_{\A\cap\B}\u} = e^{\u^\top(\logm^+\A+\logm^+\B)\u}\\
\quad&= e^{\u^\top\logm^+\A\u}\;e^{\u^\top\logm^+\B\u} 
= \tr(\A\odot\u\u^\top)\;\tr(\B\odot\u\u^\top).
\end{align*}
Needless to say Property \ref{i:prodpinch} does not
hold if $\u\u^\top$ is replaced by a mixture of dyads.

Property \ref{i:pinchinverse}. Trivially follows from \ref{i:rpinch}.
\end{proof}

We will now discuss some of the properties further. In particular, we will show a simple example
that demonstrates that the upper bound \ref{i:ubound} can
be quite loose when both matrices are dyads. 
In this case the inequality becomes:
\[\tr(\u\u^{\top}\odot\v\v^{\top})
\leq \tr(\u\u^{\top}\v\v^{\top}) = (\u\cdot\v)^2.\]
The right hand side can be made arbitrarily close to one by choosing almost
parallel $\u$ and $\v$. The left side is zero in this case, which can be seen by analyzing the intersection
of the ranges. Dyads are rank one matrices and their
ranges are lines through the origin. The intersection of two such lines is either only the origin or the line 
itself. Thus, by Property \eqref{i:inters} it follows that 
$\u\u^{\top}\odot\v\v^{\top} = \zero$, 
unless $\u=\pm\v$. 
This can also be seen from the limit expression in Equation \eqref{e:odotlim}:
\begin{eqnarray*}
\u\u^{\top}\odot\v\v^{\top} &=& \underset{n\rightarrow\infty}{\lim} 
((\u\u^{\top})^{\frac{1}{n}}(\v\v^{\top})^{\frac{1}{n}})^n  = 
\underset{n\rightarrow\infty}{\lim} (\u\u^{\top}\v\v^{\top})^n\\
 &=& \bigl(\underset{n\rightarrow\infty}{\lim}
(\u\cdot\v)^{2n-1}\bigr)\, \u\v^{\top} = \zero,\;
\text{unless $\u=\pm\v$.}
\end{eqnarray*}
Where the last equality holds because $\lvert\u\cdot\v\rvert < 1$, 
when $\u \neq \pm \v$.

Note that the expression \eqref{e:odotfancy} 
for $\odot$ based on $\logm^+$
gives us a convenient method for computing the operation
even when the matrices have some zero eigenvalues. The modified matrix logarithm $\logm^+$ is easily 
computed via Equation \eqref{e:logplus}. The matrix $\RR_\A$ containing the orthonormal basis for range of 
$\A$ can be computed using Gram-Schmidt orthogonalization procedure or 
the QR-decomposition. To compute the
basis for the intersection of $\range(\A)$ and $\range(\B)$, 
we express the intersection i.t.o. the union
and the orthogonal complement $^\perp$ of a space:
\[\range(\A) \cap \range(\B) = \left(\range(\A)^{\perp}\cup\range(\B)^{\perp}\right)^{\perp}.\]
For any matrix $\E$, an orthonormal basis for 
$\range(\E)^{\perp}$ can be
obtained by completing an orthonormal basis for $\range(\E)$ to
an orthonormal basis for the whole space. The additional basis
vectors needed are the basis for $\range(\E)^{\perp}$.
Also, if we have two matrices $\E$ and $\F$, we can get
the range for the union of their ranges just by putting all columns of $\E$ and $\F$ together into
a bigger matrix $\G = \left(\E ,\F\right)$. 
Clearly, $\range(\G) = \range(\E) \cup \range(\F)$.
Piecing all of this together gives an implementation of the $\odot$ operation.

\section{Joint Distributions}
\label{s:joint}
A density matrix defines a generalized probability distribution 
over the dyads from one space.
However we need to consider several spaces and
joint distributions over them. In the conventional case
$A,B$ denote finite sets 
$\{a_1,\ldots,a_{n_A}\}$, $\{b_1,\ldots,b_{n_B}\},$
$(P(a_i)),(P(b_j))$ 
probability vectors over these sets and $(P(a_i,b_j))$ 
is an $n_A\times n_B$
dimensional matrix of probabilities for the tuple set $A\times B$. 
In the generalized case, $\AS, \BS$
denote real finite dimensional vector spaces
of dimension $n_\AS, n_\BS$
and $\D(\AS), \D(\BS)$ are the density matrices
defining the generalized probability distributions over
these spaces. The {\em joint space} $(\AS,\BS)$ is the
tensor
product%
\footnote{%
See \cite{\bhatia} for a formal definition of tensor product between vector spaces.
For us, the tensor product of $\R^{n_\AS}$ and $\R^{n_\BS}$ is
$\R^{n_\AS n_\BS}$.} 
between the spaces $\AS$ and $\BS$, which is of
dimension $n_{\AS}n_{\BS}$. The joint distribution 
is specified by a density matrix over this joint space, denoted
by $\D(\AS,\BS)$.

We let $\D(\a),\D(\b)$ 
denote the probabilities assigned to dyads
$\a\a^\top,\b\b^\top$ from the spaces $\AS,\BS$
by the density matrices $\D(\AS),\D(\BS)$, 
respectively: 
\begin{equation}
\D(\a) := \tr(\D(\AS)\a\a^{\top}), \quad\quad
\D(\b) := \tr(\D(\BS)\b\b^{\top}).
\tag{MJ1}
\end{equation}
The conventional probability distributions 
can be seen as diagonal density matrices. A probability
distribution $(P(a_i))$ on the set $A$ is the density matrix
$\diag((P(a_i)))$. Also $P(a_j)=\e_j^\top\diag((P(a_i)))\e_j$.

To introduce the joint probability $\D(\a, \b)$ we need the Kronecker matrix product. Given 
two matrices $\E$ and $\F$ with dimensions $n\times m$ and
$p\times q$, their Kronecker product (also known as direct product or tensor product) 
$\E \otimes \F$ is a matrix with dimensions $np\times mq$ which in block form
is given as:
\[
\begin{pmatrix}
e_{11}\F& e_{12}\F& \dotsc& e_{1m}\F\\
e_{21}\F& e_{22}\F& \dotsc& e_{2m}\F\\
\hdotsfor{4}\\
e_{n1}\F& e_{n2}\F& \dotsc& e_{nm}\F\\
\end{pmatrix}
.
\]
The Kronecker product has the following useful properties:
\begin{enumerate}[label=KP\arabic*.,ref=KP\arabic*]
\item\label{kp:top} $(\E \otimes \F)^{\top} = \E^{\top} \otimes
\F^{\top}$.
\item\label{kp:prod} $(\E \otimes \F)(\G \otimes \H) = \E\G \otimes \F\H$ if the 
dimensions are appropriate.
\item\label{kp:trace} $\tr(\E \otimes \F) = \tr(\E)\tr(\F)$.
\item\label{kp:eig} If symmetric matrix $\S$ has eigenvalues $\sigma_i$ and eigenvectors $\s_i$ 
and symmetric matrix $\T$ has eigenvalues
$\tau_j$ and eigenvectors $\t_j$, 
then $\S \otimes \T$ has eigenvalues $\sigma_i\tau_j$ and
eigenvectors $\s_i \otimes \t_j$.
\item\label{kp:odot} For symmetric positive definite matrices
$\A,\B,\C,\D$, $(\A\otimes\B)\odot(\C\otimes\D) = (\A\odot\C)\otimes(\B\odot\D)$.
\end{enumerate}
The first four properties are standard.
The last property follows from the limit definition 
\eqref{e:odotlim} of the $\odot$ operation.
\begin{align*}
(\A\otimes\B)\odot(\C\otimes\D) &= \underset{n \rightarrow \infty}{\lim}
\bigl((\A\otimes\B)^{\frac{1}{n}}(\C\otimes\D)^{\frac{1}{n}}\bigr)^n\\
 &= \underset{n \rightarrow \infty}{\lim} \bigl((\A^{\frac{1}{n}}\C^{\frac{1}{n}})
\otimes(\B^{\frac{1}{n}}\D^{\frac{1}{n}}) \bigr)^n\\
 &= \bigl(\underset{n \rightarrow \infty}{\lim} (\A^{\frac{1}{n}}\B^{\frac{1}{n}})^n \bigr)\otimes
  \bigl(\underset{n \rightarrow \infty}{\lim} (\C^{\frac{1}{n}}\D^{\frac{1}{n}})^n \bigr)
\end{align*}
The last transition which moved the limit inside the Kronecker product, follows from the fact that the 
elements of the Kronecker product matrix are just pairwise products of elements from the two matrices.
And when all limits exist, a limit of a product of two number sequences is a product of limits.

Now the joint probability $\D(\a,\b)$ becomes the probability assigned by density matrix $\D(\AS,\BS)$
to the jointly specified dyad
$(\a\otimes\b)(\a\otimes\b)^\top$:
\begin{equation}
\D(\a,\b) := \tr(\D(\AS,\BS)(\a\otimes\b)(\a\otimes\b)^\top)
=\tr(\D(\AS,\BS)(\a\a^{\top}\otimes\b\b^{\top})).
\tag{MJ3}
\end{equation}
Note that in the conventional case a joint
between two sets $A$ and $B$ is defined over
all pairs of points from $A$ and $B$.
However, in the generalized case,
there are elementary events in the joint space that
don't decompose into elementary events of the marginal
density matrices, i.e.
there are dyads in the joint space that are not
of the form $(\a\otimes\b)(\a\otimes\b)^\top$. 
This is what quantum physicists call ``entanglement''.

\section{Marginalization of the Joint via Partial Traces}
\label{s:marginal}
We would like to be able to perform marginalization
operations on our joint density matrix $\D(\AS,\BS)$,
i.e. obtain the density matrix $\D(\AS)$ from the joint matrix. In
the conventional case the marginalization was performed 
by summing out one of the variables by summing the rows or the columns of the matrix specifying the
joint probability distribution. For density matrices,
the analogous operation is the {\em partial trace} (see e.g. \cite{NieChu00}). 

The partial trace is a generalization of normal matrix trace. 
It typically produces a matrix instead of a number
and can be used to retrieve
the (scaled) factor matrices from a Kronecker product.
We denote the partial trace with $\tr_{\AS}$, 
where $\AS$ specifies the space to be ``summed out''. 
Suppose $\G$ is a matrix over the space $\AS\otimes\BS$
and $\AS$ has dimension $n$ and $\BS$ dimension $m$.
Thus $\G$ has dimension $nm\times nm$ and can be written in block 
form as a $n\times n$ matrix
of $m\times m$ matrices $\G_{ij}$:
\[
\G = 
\begin{pmatrix}
\G_{11}& \G_{12}& \dotsc& \G_{1n}\\
\G_{21}& \G_{22}& \dotsc& \G_{2n}\\
\hdotsfor{4}\\
\G_{n1}& \G_{n2}& \dotsc& \G_{nn}\\
\end{pmatrix}
\]
Here we suppose that space $\AS$ is $\R^n$ and space $\BS$ is $\R^m$.
Then the two partial traces of this matrix are given by:
\[\underbrace{\tr_{\AS}(\G)}_{m \times m} = \G_{11} + \G_{22} + \dotsc + \G_{nn} \]
\[ \underbrace{\tr_{\BS}(\G)}_{n \times n} = 
\begin{pmatrix}
\tr(\G_{11})& \tr(\G_{12})& \dotsc& \tr(\G_{1n})\\
\tr(\G_{21})& \tr(\G_{22})& \dotsc& \tr(\G_{2n})\\
\hdotsfor{4}\\
\tr(\G_{n1})& \tr(\G_{n2})& \dotsc& \tr(\G_{nn})\\
\end{pmatrix}
\]
In multilinear algebra partial traces are known as tensor contractions and can of course be generalized
to the tensor product of more than two spaces.
The partial trace is a linear operator
and we now give some other useful properties:
\begin{enumerate}[label=PT\arabic*.,ref=PT\arabic*]
\item\label{pt:1} \(\tr_{\AS}(\E\otimes \F)=\tr(\E)\F,
\quad \tr_{\BS}(\E\otimes \F) = \tr(\F)\E\).
\item\label{pt:2} \(\tr(\G) = \tr(\tr_{\AS}(\G)) =
\tr(\tr_{\BS}(\G)) \).
\item\label{pt:3} \(\tr_{\AS}(\G(\I_{\AS}\otimes \F)) = \tr_{\AS}(\G)\F, 
\quad \tr_{\AS}((\I_{\AS}\otimes \F)\G) = \F\tr_{\AS}(\G)
\).
\item\label{pt:4} \(\tr(\G(\E\otimes \F)) =
\tr(\tr_{\BS}(\G(\I_{\AS}\otimes \F))\E) \).
\end{enumerate}
The first three properties are straightforward and
the last one follows from the others as follows:
\begin{equation*} 
\begin{split}
\tr(\G(\E\otimes \F)) \overset{\ref{kp:prod}}{=}&
\,\tr(\G (\I_{\AS}\otimes \F) (\E\otimes\I_{\BS})) =
\tr((\E\otimes\I_{\BS})\G(\I_{\AS}\otimes \F)) \\
\overset{\ref{pt:2}}{=}&\tr(\tr_{\BS}((\E\otimes\I_{\BS})\G(\I_{\AS}\otimes \F)))
\overset{\ref{pt:3}}{=}\,
\tr(\E\,\tr_{\BS}(\G(\I_{\AS}\otimes \F)).
\end{split}
\end{equation*}

We use the partial trace to define marginals as follows:
\begin{equation}
\D(\AS) := \tr_{\BS}(\D(\AS,\BS)), \quad  \D(\BS) := \tr_{\AS}(\D(\AS,\BS))
\tag{MJ2}
\end{equation}
The following lemma shows that $\D(\AS)$ and $\D(\BS)$
defined this way are again density matrices.
\begin{lemma}
\label{l:partdens}
Partial trace of a density matrix is also a density matrix.
\end{lemma}
\begin{proof}
Symmetry is obvious. Trace one follows from Property \ref{pt:2} of the partial trace:
\[\tr(\D(\AS)) = \tr(\tr_B(\D(\AS,\BS))) = \tr(\D(\AS,\BS)) = 1 .\]
Positive definiteness follows by a similar argument:
\begin{eqnarray*}
\a^{\top}\D(\AS)\a &=& \tr(\D(\AS)\a\a^{\top}) 
\overset{\ref{pt:3}}{=}
\tr(\tr_B(\D(\AS,\BS)(\a\a^{\top}\otimes\I_{\BS})))\\
&\overset{\ref{pt:2}}{=}& \tr(\D(\AS,\BS)(\a\a^{\top}\otimes
\underbrace{\sum_i \b_i\b_i^{\top}}_{\I_{\BS}}))\\
&=& \sum_i \tr(\D(\AS,\BS)(\a\a^{\top} \otimes \b_i^{}\b_i^{\top}))\\
&=& \sum_i (\a\otimes \b_i)^{\top}\D(\AS,\BS)(\a\otimes
\b_i) \geq 0.
\end{eqnarray*}
\end{proof}
Partial traces also allow us to define objects of the type $\D(\AS,\b)$.
In the conventional case this corresponds to taking one row or column
out of the joint probability table. In the generalized case 
we want the following property to be satisfied: 
\begin{equation}
\label{e:margprop}
\tr(\D(\AS,\b)\a\a^{\top}) 
= \D(\a,\b).
\tag{MJ5}
\end{equation}
This is accomplished by defining $\D(\AS,\b)$ via the following formula:
\begin{equation}
\D(\AS,\b) := \tr_{\BS}(\D(\AS,\BS)(\I_A\otimes \b\b^{\top})).
\tag{MJ4}
\end{equation} 
Property MJ5 now follows from partial trace Property \ref{pt:4}. We can also see that trace of $\D(\A,\b)$
gives us the probability $\D(\b)$:
\begin{eqnarray}
\label{e:margtrace}
\tr(\D(\AS,\b)) &\overset{\text{MJ4}}{=}& \tr(\tr_{\BS}(\D(\AS,\BS)(\I_A\otimes \b\b^{\top}))) 
\overset{\ref{pt:2}}{=} \tr(\tr_{\AS}(\D(\AS,\BS)(\I_A\otimes \b\b^{\top})))\\
\nonumber
&\overset{\ref{pt:3}}{=}&
\tr(\tr_{\AS}(\D(\AS,\BS))\b\b^\top) \overset{\text{MJ2}}{=} 
\tr(\D(\BS)\b\b^\top) = \D(\b).
\end{eqnarray}
A brief note on matrix properties of $\D(\AS,\b)$. We just saw that its trace is $\D(\b)$ which
is between zero and one. Since it satisfies Property \eqref{e:margprop}, it is positive definite as well. 
Symmetry is also easily verified.

Note that for any 
orthogonal system $\b_i$ of $\BS$, 
$$\D(\AS) = \sum_i \D(\AS,\b_i).$$
This can be seen as follows.
\begin{eqnarray*}
\D(\AS) &\overset{\text{MJ2}}{=}& \tr_\BS(\D(\AS,\BS)) 
= \tr_\BS(\D(\AS,\BS)(\I_\AS\otimes\I_\BS))\\
&=& \tr_\BS(\D(\AS,\BS)(\I_\AS\otimes\underbrace{\sum_i \b_i\b_i^\top}_{\I_\BS})) 
= \sum_i \tr_\BS(\D(\AS,\BS)(\I_\AS\otimes\b_i\b_i^\top)) 
\overset{\text{MJ4}}{=} \sum_i \D(\AS, \b_i).
\end{eqnarray*}
The conventional definition of independence also naturally generalizes: $\D(\AS)$ is independent of $\D(\BS)$ 
if the joint density 
matrix decomposes: $\D(\AS,\BS) = \D(\AS) \otimes \D(\BS)$. 
It is easy to see that in this case we have $\D(\a,\b) =
\D(\a)\D(\b)$ for all $\a,\b$: 
\begin{equation*}
\begin{split}
\D(\a,\b) =&\; \tr((\D(\AS)\otimes\D(\BS))(\a\a^\top\otimes\b\b^\top)) 
\overset{\ref{kp:prod}}{=} \tr((\D(\AS)\a\a^\top)\otimes(\D(\BS)\b\b^\top))\\
\overset{\ref{kp:trace}}{=}&\;
\tr(\D(\AS)\a\a^\top)\tr(\D(\BS)\b\b^\top) = \D(\a)\D(\b).
\end{split}
\end{equation*}

\section{Conditional Probabilities}
\label{s:cond}
The topic of conditional probabilities 
in this generalized setting contains many subtleties. 
First we will give the defining
formulas for conditional density matrices and then
discuss some of the issues.
\begin{enumerate}[label=CP\arabic*.,ref=CP\arabic*]
\item $\D(\AS|\BS) := \D(\AS,\BS)\odot(\I_{\AS}\otimes\D(\BS))^{-1}$
\label{cp:A.B}
(Formula (4) of \cite{ca-qecp-99}
expressed with the $\odot$ operation). This formula requires $\D(\BS)$ to be invertible.
In the conventional case, this corresponds to the conditional probabilities being undefined if the event
conditioned on has probability zero.

\item $\D(\AS|\b) := \frac{\D(\AS,\b)}{\D(\b)}$.
\label{cp:A.b}

\item $\D(\a|\BS) := \D(\a,\BS)\odot\D(\BS)^{-1}$.
\label{cp:a.B}

\item $\D(\a|\b) := \frac{\D(\a,\b)}{\D(\b)}$. 
\label{cp:a.b}
This basic conditional probability is 
a straightforward generalization of the conventional case.
It also has a quantum-mechanical interpretation. See Appendix \ref{a:condmeas} for details.
\end{enumerate}

Note that \ref{cp:A.B} has the form:
density matrix $\odot$ inverse of a normalization.
We can also reexpress the other definitions in this unified form:

\begin{enumerate}[label=CP$'$\arabic*.,ref=CP$'$\arabic*,start=2]
\item $\displaystyle\D(\AS|\b) = \underbrace{\tr_\BS(\D(\AS,\BS)(\I_\AS\otimes\b\b^{\top}))}_{\D(\AS,\b)}
\odot\underbrace{\tr_\BS((\I_\AS\otimes\D(\BS))(\I_\AS\otimes\b\b^{\top}))^{-1}}_{\frac{1}{\D(\b)}\I_{\AS}}$.

\item $\D(\a|\BS) = \underbrace{\tr_\AS(\D(\AS,\BS)(\a\a^{\top}\otimes\I_\BS))}_{\D(\a,\BS)}
\odot\underbrace{\tr_\AS((\I_\AS\otimes\D(\BS))(\a\a^{\top}\otimes\I_\BS))^{-1}}_{\D(\BS)^{-1}}$.

\item $\displaystyle\D(\a|\b) = \underbrace{\tr(\D(\AS,\BS)(\a\a^{\top}\otimes\b\b^{\top}))}_{\D(\a,\b)}
\odot
\underbrace{\tr((\I_\AS\otimes\D(\BS))(\a\a^{\top}\otimes\b\b^{\top}))^{-1}}_{\frac{1}{\D(\b)}}$.
\end{enumerate}

We say that the joint density $\D(\AS,\BS)$
is \textit{decoupled} if its eigendecomposition has
the form: $\D(\AS, \BS) 
= (\WW_\AS\otimes\WW_\BS)\diag(\ww)(\WW_\AS\otimes\WW_\BS)^\top$. 
Note that $\WW_\AS\otimes\WW_\BS$ is orthogonal iff
both $\WW_\AS$ and $\WW_\BS$ are orthogonal.
As we shall see later, 
dealing with conditionals is often simpler in the decoupled case.
We first prove an upper bound for $\tr(\D(\AS|\BS))$ that
is tight iff the joint is decoupled.
\begin{lemma}
\label{l:decoupled}
The following inequality holds:
\[\tr(\D(\AS|\BS)) \leq n_{\BS}, \]
where $n_\BS$ is the dimensionality of space $\BS$. Furthermore, $\tr(\D(\AS|\BS)) = n_{\BS}$ 
if and only if the joint $\D(\AS,\BS)$ is decoupled.
\end{lemma}
\begin{proof}
The inequality is shown using properties of $\odot$ and partial traces:
\begin{eqnarray*}
\tr(\D(\AS|\BS)) &\overset{\ref{cp:A.B}}{=}& \tr(\D(\AS,\BS)\odot(\I_{\AS}\otimes\D(\BS)^{-1}))
\overset{\ref{i:ubound}}{\leq} \tr(\D(\AS,\BS)(\I_{\AS}\otimes\D(\BS)^{-1}))\\
&\overset{\ref{pt:2}}{=}& \tr(\tr_{\AS}(\D(\AS,\BS)(\I_{\AS}\otimes\D(\BS)^{-1})))
\overset{\ref{pt:3}}{=} 
\tr(\underbrace{\tr_{\AS}(\D(\AS,\BS))}_{\D(\BS)} \D(\BS)^{-1}) \\
&=& \tr(\I_{\BS}) = n_{\BS}.
\end{eqnarray*}
Remember that equality in Property \ref{i:ubound} of $\odot$ only occurs when the two matrices commute. 
Two matrices commute iff their eigensystems are the same. This gives us the condition that the eigensystem
of $\D(\AS,\BS)$ must be the same as the eigensystem of
$\I_{\AS}\otimes\D(\BS)^{-1}$. The latter 
eigensystem is clearly decoupled. Thus for equality to hold it is \textit{necessary}
that the eigensystem of $\D(\AS, \BS)$ be decoupled. 

Now we will argue that it is also \textit{sufficient}.
Let the joint density matrix have eigensystem 
$\D(\AS,\BS) = (\WW_{\AS}\otimes\WW_{\BS})\diag(\boldsymbol{\omega})
(\WW_{\AS}^\top\otimes\WW_{\BS}^\top)$.
$\I_{\AS}$ commutes with any matrix on space $\AS$. Therefore it suffices to show that the marginal
$\D(\BS)$ in this case has eigensystem $\WW_{\BS}$. The decoupled eigensystem matrix
$\WW_{\AS}\otimes\WW_{\BS}$ has the following list of $n_{\AS}n_{\BS}$ colums:
\begin{eqnarray*}
\WW_{\AS}\otimes\WW_{\BS} \;\;= 
&\Bigl(\w_{\AS}^1\otimes\w_{\BS}^1,\w_{\AS}^1\otimes\w_{\BS}^2, \dotsc , 
\w_{\AS}^1\otimes\w_{\BS}^{n_{\BS}},\\ 
&\;\;\w_{\AS}^2\otimes\w_{\BS}^1, \w_{\AS}^2\otimes\w_{\BS}^2, \dotsc, \w_{\AS}^2\otimes\w_{\BS}^{n_\BS},\\
&\;\;\dotsc,\dotsc,\dotsc,\dotsc,\dotsc,\dotsc,\dotsc,\dotsc,\dotsc\\
&\;\;\;\;\;\;\;\;\w_{\AS}^{n_{\AS}}\otimes\w_{\BS}^{1},\w_{\AS}^{n_{\AS}}\otimes\w_{\BS}^{2}
\dotsc,\w_{\AS}^{n_{\AS}}\otimes\w_{\BS}^{n_{\BS}} \Bigr).
\end{eqnarray*}
In correspondence with this structure
we adopt a double indexing scheme for the eigenvalues $\omega_{i,j}$ of the joint matrix 
$\D(\AS,\BS)$, where $\omega_{i,j}$ is the eigenvalue
associated with eigenvector $\w^i_{\AS}\otimes \w^j_{\BS}$.
The index $i$ runs from $1$ to $n_{\AS}$, and $j$ runs to $n_{\BS}$. Now the
eigendecomposition can be written as:
\[\D(\AS,\BS) = \sum_{i,j} \omega_{i,j}\; 
(\w_{\AS}^i(\w_{\AS}^i)^\top\otimes \w_{\BS}^j(\w_{\BS}^j)^\top). \]
Partial trace is a linear operator and 
$\tr_{\AS}(\w_{\AS}^i(\w_{\AS}^i)^\top\otimes \w_{\BS}^j(\w_{\BS}^j)^\top) 
\overset{\ref{pt:1}}{=} \w_{\BS}^j(\w_{\BS}^j)^\top$. 
Therefore:
\[\D(\BS) = \tr_{\AS}(\D(\AS,\BS)) = \sum_j \omega_j\; \w_{\BS}^j(\w_{\BS}^j)^\top,\]
where $\omega_j = \sum_{i} \omega_{i,j}$. Thus we produced the eigendecomposition of the
marginal $\D(\BS)$ and it indeed has eigensystem $\WW_{\BS}$.
\end{proof}

Let us briefly discuss the connection and difference between our notion of decoupled joints and the notion
of entanglement that appears in quantum physics. Recall that entanglement, as we mentioned at the
end of Section \ref{s:joint} corresponds to the fact that there are dyads $\c\c^\top$ in the joint space
$(\AS,\BS)$ that can't be written as $\a\a^\top\otimes\b\b^\top$ for any two dyads $\a\a^\top$ and 
$\b\b^\top$ in $\AS$ and $\BS$. This notion carries over to mixed states or density matrices. In quantum
physics, a joint density matrix $\D(\AS,\BS)$ is called \textit{separable} (or non-entangled) 
if it can be expressed as
$(\WW_\AS\otimes\WW_\BS)\diag(\ww)(\WW_\AS\otimes\WW_\BS)^\top$. 
The crucial difference between the
definitions of separable and decoupled matrices is 
that in the separable case, $\WW_\AS$ and $\WW_\BS$ don't
have to be orthogonal.
Every decoupled matrix is separable, but there are separable
density matrices that are not decoupled.
The question of deciding whether a given matrix is separable is known to be very difficult, whereas
the question of being decoupled is easily decided by e.g.
the condition of the above lemma.
One of the reasons for which \cite{ca-qecp-99} introduced a
conditional density matrix via Rule \ref{cp:A.B} was to give a 
necessary condition for the separability of a joint density matrix.

To complete the rules for conditional density matrices, we would need rules that allow us to marginalize the
conditionals, e.g. for going from $\D(\AS|\BS)$ to $\D(\a|\b)$.
One obvious consequence of our definitions is the 
marginalization rule for $\D(\AS|\b)$:
\begin{equation}
\D(\a|\b) 
\overset{\ref{cp:a.b}}{=} \frac{\D(\a,\b)}{\D(\b)}
\overset{\text{MJ5}}{=} \frac{\tr(\D(\A,\b)\a\a^\top)}{\D(\b)}
\overset{\ref{cp:A.b}}{=} \tr(\D(\AS|\b)\a\a^{\top}). 
\tag{MC4}
\end{equation}
There don't seem to be any other
simple marginalization rules for $\D(\AS|\BS)$ and $\D(\a|\BS)$
that hold for arbitrary joints.
However, when the joint is decoupled,
then the following additional marginalization rule
for $\D(\AS|\BS)$ is valid:
\begin{lemma}
For all decoupled joints $\D(\AS,\BS)$,
\[\D(\a|\BS) = \tr_{\AS}(\D(\AS|\BS)(\a\a^\top\otimes\I_\BS)).\]
\end{lemma}
\begin{proof}
We will compute both sides of the equation and show them to be identical. 
We begin by writing down the decomposition of the decoupled joint from Lemma \ref{l:decoupled}:
\begin{equation}
\label{e:decoupled}
\D(\AS,\BS) = \sum_{i,j} \omega_{i,j}\; 
(\w_{\AS}^i(\w_{\AS}^i)^\top\otimes \w_{\BS}^j(\w_{\BS}^j)^\top).
\end{equation}
Additionally, in the same lemma, the following form for $\D(\BS)$ was established in this case:
\begin{equation}
\label{e:bdecoupled}
\D(\BS) = \tr_{\AS}(\D(\AS,\BS)) = \sum_j \omega_j\; \w_{\BS}^j(\w_{\BS}^j)^\top,
\end{equation}
where $\omega_j = \sum_{i} \omega_{i,j}$. According to \ref{cp:a.B}, 
$\D(\a|\BS) = \D(\a,\BS)\odot\D(\BS)^{-1}$, therefore we will need to compute $\D(\a,\BS)$:
\begin{equation*}
\begin{split}
\D(\a,\BS) \overset{\text{MJ4}}{=}& \tr_{\AS}(\D(\AS,\BS)(\a\a^\top \otimes \I_\BS))
\overset{\eqref{e:decoupled}}{=} 
\sum_{i,j} \omega_{i,j}\;\tr_{\AS}((\w_{\AS}^i(\w_{\AS}^i)^\top\otimes \w_{\BS}^j(\w_{\BS}^j)^\top)
(\a\a^\top \otimes \I_\BS))\\
\overset{\ref{kp:prod}}{=}& \sum_{i,j} \omega_{i,j}\;
\tr_{\AS}((\w_{\AS}^i(\w_{\AS}^i)^\top)\a\a^\top \otimes \w_{\BS}^j(\w_{\BS}^j)^\top)
\overset{\ref{pt:1}}{=} \sum_{i,j} \omega_{i,j}(\w_{\AS}^i\cdot\a)^2\; \w_{\BS}^j(\w_{\BS}^j)^\top.
\end{split}
\end{equation*}
Together with \eqref{e:bdecoupled}, this gives:
\[\D(\a|\BS) = \sum_{i,j} \frac{\omega_{i,j}(\w_{\AS}^i\cdot\a)^2}{\omega_j}
\; \w_{\BS}^j(\w_{\BS}^j)^\top \]

Now, we proceed to the right side of the equation in the lemma. 
Substituting \eqref{e:decoupled} and \eqref{e:bdecoupled} into the formula for $\D(\AS|\BS)$ 
we obtain:
\begin{equation}
\label{e:conddecoupled}
\D(\AS|\BS) \overset{\ref{cp:A.B}}{=} \D(\AS,\BS)\odot(\I_\AS\otimes\D(\BS)^{-1})
= \sum_{i,j} \frac{\omega_{i,j}}{\omega_j}\,(\w_{\AS}^i(\w_{\AS}^i)^\top\otimes \w_{\BS}^j(\w_{\BS}^j)^\top).
\end{equation}
Using linearity of the partial trace we compute the right side as follows:
\begin{equation*}
\begin{split}
\tr_{\AS}(\D(\A|\B)(\a\a^\top\otimes\I_{\BS})) \overset{\eqref{e:conddecoupled}}{=}&
\sum_{i,j} \frac{\omega_{i,j}}{\omega_j}\; \tr_{\AS}(
(\w_{\AS}^i(\w_{\AS}^i)^\top\otimes \w_{\BS}^j(\w_{\BS}^j)^\top)
(\a\a^\top\otimes\I_{\BS}))\\
\overset{\ref{kp:prod}}{=}& \sum_{i,j} \frac{\omega_{i,j}}{\omega_j}\;
\tr_{\AS}((\w_{\AS}^i(\w_{\AS}^i)^\top)\a\a^\top \otimes \w_{\BS}^j(\w_{\BS}^j)^\top)\\
\overset{\ref{pt:1}}{=}&
\sum_{i,j} \frac{\omega_{i,j}(\w_{\AS}^i\cdot\a)^2}{\omega_j}
\; \w_{\BS}^j(\w_{\BS}^j)^\top.
\end{split}
\end{equation*}
\end{proof}
As discussed, obtaining 
$\D(\a|\b)
=\frac{\tr(\D(\AS,\BS)(\a\a^\top\otimes\b\b^\top))}
      {\tr(\D(\BS)\b\b^\top)}$
from $\D(\AS|\BS)$ is non-trivial. 
In particular, there are cases where
\[
\tr(\D(\AS|\BS)(\a\a^\top\otimes\b\b^\top)) \neq D(\a|\b),
\]
even when $\D(\AS,\BS)$ is decoupled and $\a$ and $\b$
are not eigenvectors of $\D(\AS)$ and $\D(\BS)$,
respectively.
Curiously enough, 
if we replace the matrix product with $\odot$, then 
we always have
\begin{align*}
&\tr(\D(\AS|\BS)\odot(\a\a^\top\otimes\b\b^\top))\\
&\quad \overset{\ref{cp:A.B}}{=} 
\tr((\D(\AS,\BS)\odot(\I_{\AS}\otimes\D(\BS)^{-1}))\odot(\a\a^\top\otimes\b\b^\top))\\
&\quad\overset{\ref{i:prodpinch}}{=} \tr(\D(\AS,\BS)\odot(\a\a^\top\otimes\b\b^\top))
\;\;\tr((\I_\AS\otimes\D(\BS)^{-1})\odot(\a\a^\top\otimes\b\b^\top))\\
&\quad\overset{\ref{i:pinchinverse}}{=} 
\frac{\tr(\D(\AS,\BS)\odot(\a\a^\top\otimes\b\b^\top))}{\tr(\D(\BS)
\odot \b\b^\top)}.
\end{align*}

Let us now recall the conditionals in the conventional
probability theory. The full conditional table $P(A|B)$
lists conditional probabilities of all pairs of elementary events $P(a_i|b_j)$. This table has the obvious
properties: The sum of all entries is $n_B$ 
and the sum of any column is 1, i.e. $\sum_i P(a_i|b_j) = 
\sum_i \frac{P(a_i,b_j)}{P(b_j)} = \frac{P(b_j)}{P(b_j)} = 1$. Thus a conditional table is a column-stochastic
matrix and for any such matrix 
we can construct a joint that has that matrix as its conditional
table. For example we can take arbitrary probability vector $p$ and multiply the $i$-th
column of $P(A|B)$ by $p_i$, now the sum of each column is $p_i$ and thus the sum of all entries is $1$ and
we have a valid joint. Note that this implies that many different joints have the same conditional
table. 

The decoupled case behaves as the conventional case,
i.e. many joints correspond to the same conditional.
A decoupled joint and conditional 
always have the same eigensystem
and going from the joint to the conditional
is similar to the conventional case (See
(\ref{e:decoupled}-\ref{e:conddecoupled}) for details).

However, for non-decoupled joint density matrices, i.e. when 
$\tr(\D(\AS|\BS)) < n_{\BS}$ (Lemma \ref{l:decoupled}),
the situation is quite different.
For example,
the eigenvalues of $\D(\AS|\BS)$ can now be bigger 
than 1 \cite{ca-qecp-99}.
Also based on numerical experiments, 
we conjecture 
that in the non-decoupled case, the mapping between
$\D(\AS,\BS)$ and $\D(\AS|\BS)$ is invertible, i.e. 
unlike the conventional case, there is only one joint
that gives rise to a given conditional matrix. 
In other words we conjecture that in the non-decoupled case it
suffices to specify the conditional $\D(\AS|\BS).$

More specifically, we claim that the following 
EM-like algorithm converges to $\D(\BS)$
and then $\D(\AS,\BS) \overset{\ref{cp:A.B}}{=}\D(\AS|\BS)\odot\D(\BS)$:
$\W_0$ is initialized to $\I_\BS/n_\BS$ and
the estimate $\W_{t+1}$ for $\D(\BS)$ is computed from $\D(\AS|\BS)$ 
and the previous estimate $\W_t$ as 
\[\W_{t+1} = \frac{\tr_{\AS}(\D(\AS|\BS)\odot(\I_{\AS}\otimes
\W_t))}{\tr(\D(\AS|\BS)\odot(\I_{\AS}\otimes \W_t))}. \]

\section{Theorems of Total Probability}
\label{s:total}
The Theorem of Total Probability is an important
calculation in conventional probability theory. It expresses
probability of some event $a$ as an expected 
conditional probability of the elementary events
$b_i$ that form a partition of the probability space $B$:
\[P(a) = \sum_i  P(a|b_i)P(b_i).\]
\begin{enumerate}[label=TP\arabic*.,ref=TP\arabic*]
\item\label{tp:small}
For any 
orthogonal system $\b_i$ of $\BS$,
$\D(\a) = \sum_i \D(\a|\b_i)\D(\b_i)$.
\item \label{tp:basic}
$\D(\a) = \tr(\D(\a|\BS)\odot\D(\BS))$
\label{TP:bayes}
\item\label{tp:big}
 $\D(\AS) = \tr_{\BS}(\D(\AS|\BS)\odot(\I_{\AS}\otimes\D(\BS)))$.
\end{enumerate}
The first formula can be shown as follows: 
\[\D(\a) = \tr(\D(\a,\BS)) = \sum_i \b_i^{\top}\D(\a,\BS)
\b_i = \sum_i \D(\a,\b_i) = \sum_i \D(\a|\b_i)\D(\b_i).\]
To derive the second apply $\odot \D(\BS)$ to both sides of \ref{cp:a.B}, 
take trace of both sides and use \eqref{e:margtrace}.
The proof of the third property follows the same outline
but uses \ref{cp:A.B} and MJ2.

Conventional versions of the last two properties
are obtained when the density and conditional matrices are diagonal. 
Note that in general these generalizations of the 
Theorem of Total Probability do not ``decouple'', i.e. you cannot write them as a sum of products of
conditional and marginal probabilities. However, using the Property \ref{i:ubound} of $\odot$ operation
we can establish upper bounds on probability of $\D(\a)$ in terms of ``decoupled'' sums that look
like the conventional versions of the Theorem of Total Probability. 
If $\D(\BS) = \sum_i \omega_i\; \w_i\w_i^\top$ and $\D(\a|\BS) = \sum_i \lambda_i\; \u_i\u_i^\top$
are eigendecompositions of the corresponding matrices, then
\begin{eqnarray}
\D(\a) &=& \tr(\D(\a|\BS)\odot\D(\BS)) \leq \tr(\D(\a|\BS)\D(\BS)) 
\nonumber \\
 &=& \underbrace{\sum_i \overbrace{\omega_i}^{\text{probability}}
\overbrace{\w_i^\top\D(\a|\BS)\w_i}^{\text{variance}}}_{\text{expected variance}}
\nonumber\\
 &=& \underbrace{\sum_i \overbrace{\u_i^\top\D(\BS)\u_i}^{\text{probability $\D(\u_i)$}}
\overbrace{\lambda_i}^{\text{outcome}}}_{\text{expected measurement}}.
\label{e:expected}
\end{eqnarray}
The first version of the upper bound corresponds to using the eigendecomposition of $\D(\BS)$ and
can be interpreted as an expected variance calculation with $\D(\a|\BS)$ as the covariance matrix.
The second version expands $\D(\a|\BS)$ and corresponds to a quantum measurement of system
in state $\D(\BS)$ with instrument specified by $\D(\a|\BS)$.
Letting $p(b_i)$ equal $\omega_i$ or $\u_i^\top\D(\BS)\u_i$ and letting $p(a_i|b_i)$ equal 
$\w_i^\top\D(\a|\BS)\w_i$ or $\lambda_i$, we see the correspondence of these upper bounds to the
conventional Theorem of Total Probability. The equality only occurs when $\D(\a|\B)$ and $\D(\B)$ commute.

\section{Bayes Rules}
\label{s:bayes}
In the conventional setup we assume that a model $M_i$ is
chosen with prior probability $P(M_i)$. The model then
generates the data $y$ with probability $P(y|M_i)$, 
i.e. 
\begin{eqnarray*}
P(y)&=&\sum_i P(M_i)P(y|M_i)
\\&=&\tr(\diag\left((P(M_i)\right) 
         \diag\left((P(y|M_i))\right).
\end{eqnarray*}
The reason why we expressed $P(y)$ as a trace of two diagonal
matrices will become apparent in a moment.

The generalized setup is completely analogous.
There is an underlying joint space $(\MS,\YS)$ between
the model space $\MS$ and the data space $\YS$.
The prior is specified by a density matrix $\D(\MS)$.
The data is a unit direction $\y$ in $\YS$ space that is
generated by the density $\D(\YS)$. The probability
$\D(\y)$ can be expressed i.t.o. the prior $\D(\MS)$ and data likelihood
$\D(\y|\MS)$ using \ref{TP:bayes}:
$$\D(\y)=\tr(\D(\MS) \odot \D(\y|\MS)).$$
Note that in the conventional case we first chose
a model based on the prior and then generated
data based on the chosen model.
In the generalized case we do not know how to decouple
the action on the prior from the choice of the data when
conditioned on the prior.

Let us first recall the conventional Bayes rule
and rewrite it in matrix notation:
\begin{equation}
\label{e:bayes}
P(M_i|y)=
\frac{P(M_i)P(y|M_i)}{P(y)}, \text{ where } P(y) = \sum_j P(M_j)P(y|M_j)
\end{equation}
\[\diag\left(P(M_i|y)\right)
=
\frac{ \diag\left(P(M_i)\right)\diag\left(P(y|M_i)\right)}
{\tr\left(\diag\left(P(M_i)\right)\diag\left(P(y|M_i)\right)\right)}.
\]

We now present and discuss the analogous Bayes rule for the
generalized setting.
At the end of this section we present a list of all Bayes rules.

\begin{figure}
\begin{minipage}[b]{.45\textwidth}
\includegraphics[width=\textwidth]{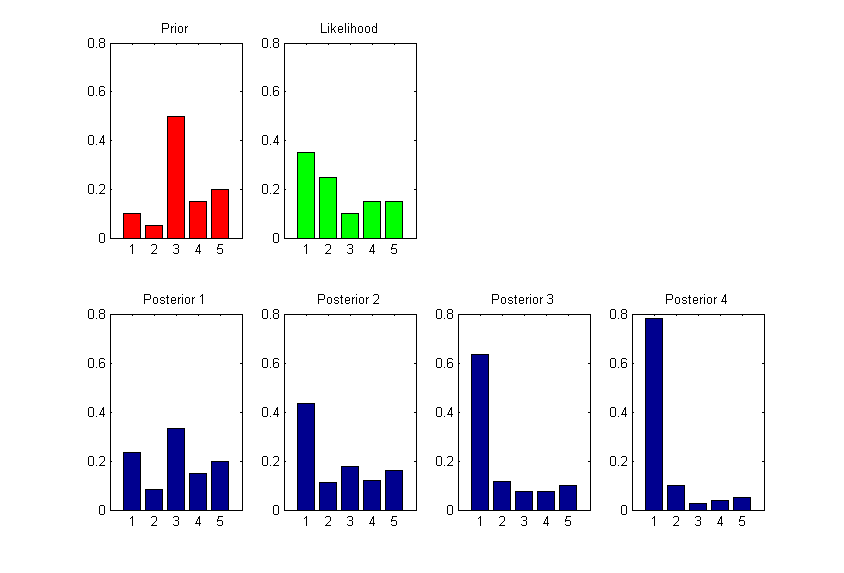}
\caption{We apply the conventional Bayes rule 4 times, using the
the same data likelihood vector $P(y|M_i)$ and making the current posterior the new prior.
At first, the posteriors are close to the initial prior
but eventually the posteriors focus their weight
on $\argmax_i P(y|M_i)$. The conventional Bayes rule may be seen as a soft maximum
calculation. The initial prior is in red, the likelihood is in green and posteriors are in blue.}
\label{f:bayes}
\end{minipage}
\hfill
\begin{minipage}[b]{.50\textwidth}
\begin{center}
\includegraphics[width=1\textwidth]{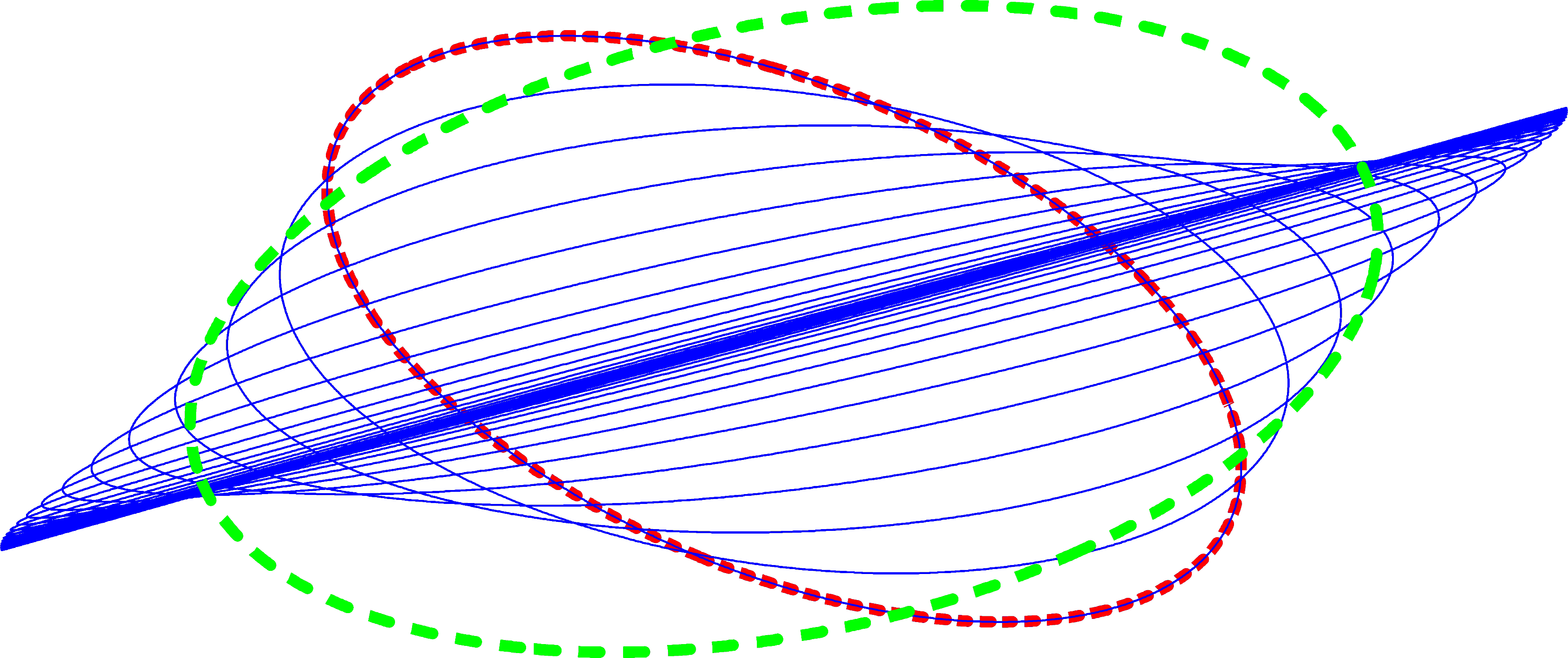}
\end{center}
\caption{We depict several iterations of the
generalized Bayes rule. The red ellipse depicts the prior
$\D(\MS)$, the green ellipse depicts the data likelihood
matrix $\D(\y|\MS)$, which is kept fixed on successive
iterations, and the blue ellipses depict posteriors
$\D(\MS|\y)$. The posterior density
matrices gradually move away from the prior and focus on
the longest axis of the covariance matrix. The generalized Bayes rule can be seen as a soft 
calculation of eigenvector with largest eigenvalue. }
\label{f:worm}
\end{minipage}\\
\begin{minipage}[t]{0.47\textwidth}
\centerline{\includegraphics[width=\textwidth]{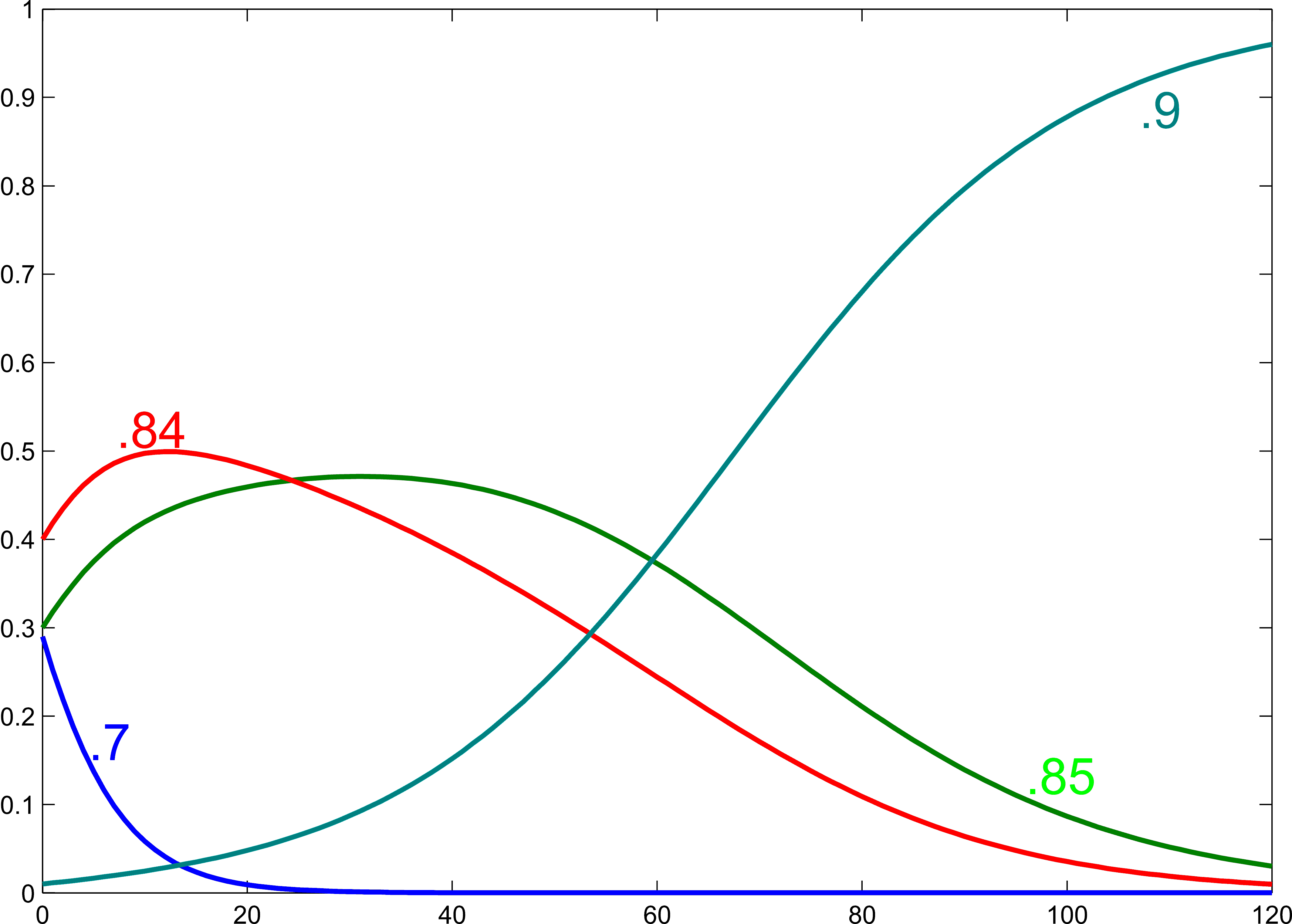}}
\caption{We plot many iterations of the conventional Bayes
rule when the same data likelihood $(P(y|M_i)) =
(.7,.84,.85,.9)$ is
used in each iteration and the prior is
$(P(M_i))=(.29,.4,.3,.01)$.
For each of the four models we plot the posterior probability as
a function of the iteration number.
Initially the posterior curve 
with likelihood .85 overtakes the curve
with likelihood .84, but eventually the curve with
likelihood .9 takes over both.
Note that the curve with the largest data likelihood looks like
a sigmoid and the one with smallest like a reverse sigmoid.}
\label{f:sigmadiag}
\end{minipage}
\hfill
\begin{minipage}[t]{0.47\textwidth}
\centerline{\includegraphics[width=\textwidth]{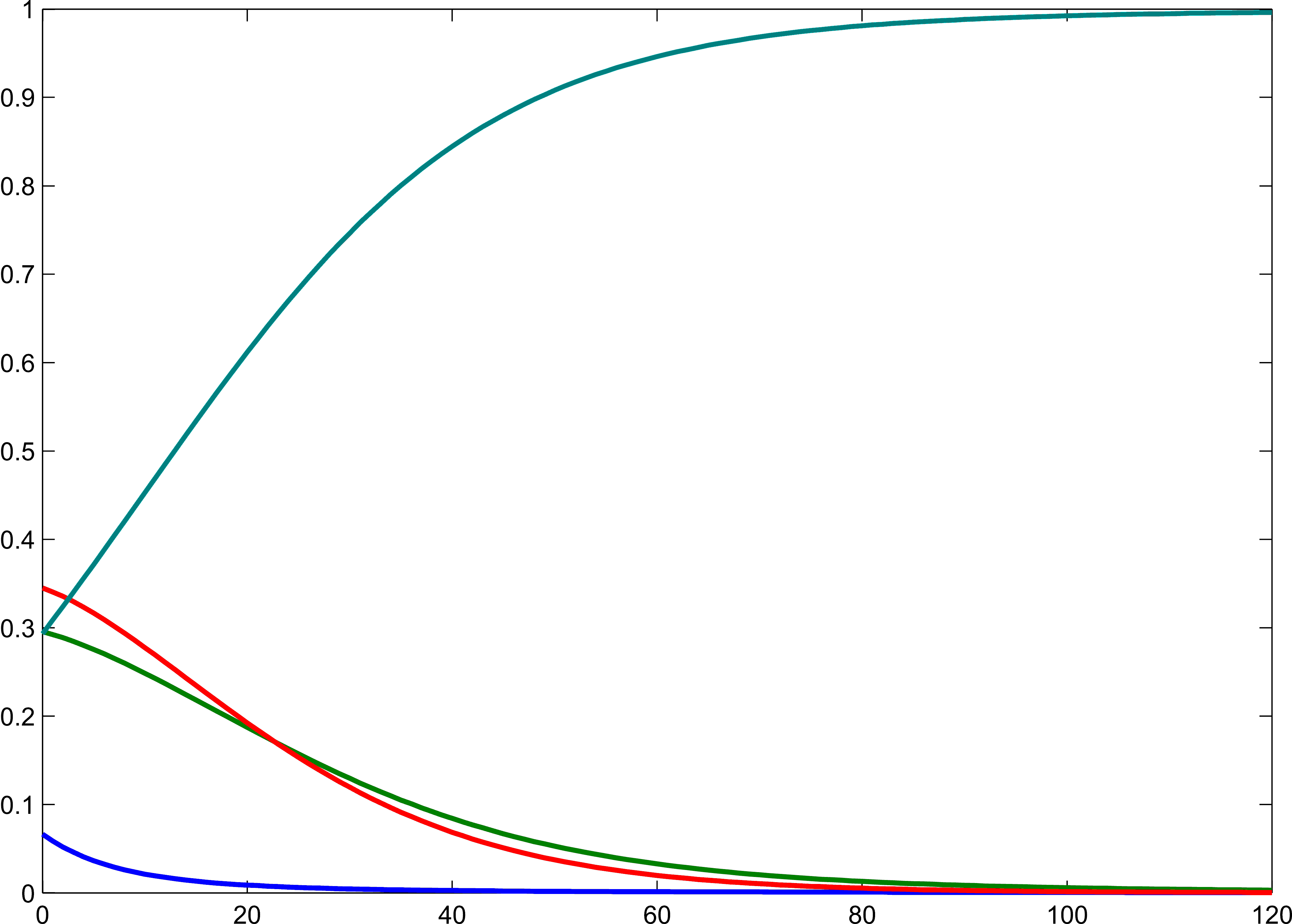}}
\caption{We plot many iterations of the generalized Bayes
rule when the same data likelihood matrix $\D(y|\MS)$ is
used in each iteration. As the prior $\D(\MS)$ we choose
the diagonalized prior $\diag((P(M_i))$ of Figure \ref{f:sigmadiag}
on the left and as the likelihood $\D(y|\MS)$ we
choose $\U \diag((P(y|M_i)) \U^T,$ 
where the eigensystem $\U$ is a random rotation matrix.
Let $\D(\MS|t)$ denote the posterior at iteration $t$
when the fixed $\D(y|\MS)$ is used in all iterations.
The curves are the projections of this posterior
onto the four eigendirections of $\D(y|\MS)$ 
as a function of $t$, 
i.e. $\u_i^\top \D(\MS|t) \u_i,$ where $\u_i$ are the
columns of $\U$. 
The above plot is qualitatively similar to the left plot.
The curve corresponding to the largest
eigenvalue of the data likelihood is again a partial sigmoid.}
\label{f:sigmarot}
\end{minipage}
\end{figure}
In the generalized Bayes rule we cannot simply
multiply the prior density matrix with the 
data likelihood matrix.
This is because
a product of two symmetric positive definite
matrices can be neither symmetric nor positive definite
(See Figure \ref{f:odot}). 
Instead, we replace the matrix multiplication with $\odot$ operation:
\begin{equation}
\label{e:gbayes}
\D(\MS|\y)
=\frac{\D(\MS) \odot \D(\y|\MS)}
{\D(\y)}, \text{ where } \D(\y) = \tr(\D(\MS)\odot\D(\y|\MS)).
\end{equation}
Normalizing by the trace ensures that the trace of the
posterior density matrix is one. 
In both the conventional as well as the new Bayes rule above,
the normalization constant is the likelihood of the data.
When the matrices $\D(\MS)$ and $\D(\y|\MS)$ have the same
eigensystem, then $\odot$ becomes the matrix multiplication.
In the following subsections we derive
the above Bayes rules from the minimum 
relative entropy principle.
For the conventional Bayes rule the standard relative entropy
between probability vectors is used,
whereas the generalized Bayes rule
and the crucial $\odot$ operation
is motivated by the quantum relative entropy between
density matrices due to Umegaki (see e.g. \cite{NieChu00}).

We visualize the conventional Bayes rule in Figure \ref{f:bayes}. Repeated application of the rule with
the same likelihood makes the posteriors increasingly concentrated on the point with maximum 
data likelihood $P(y|M_i)$. Therefore this rule can be interpreted as a soft max-likelihood calculation.
Figure \ref{f:worm} demonstrates the generalized Bayes rule.
There the posterior gradually moves towards the eigenvector
belonging to the largest eigenvalue 
of the data likelihood matrix $\D(\y|\MS)$.
Thus the new rule can be interpreted as a soft 
calculation of the eigenvector with maximum eigenvalue. 

\begin{figure}[t]
\centerline{\includegraphics[width=\textwidth]{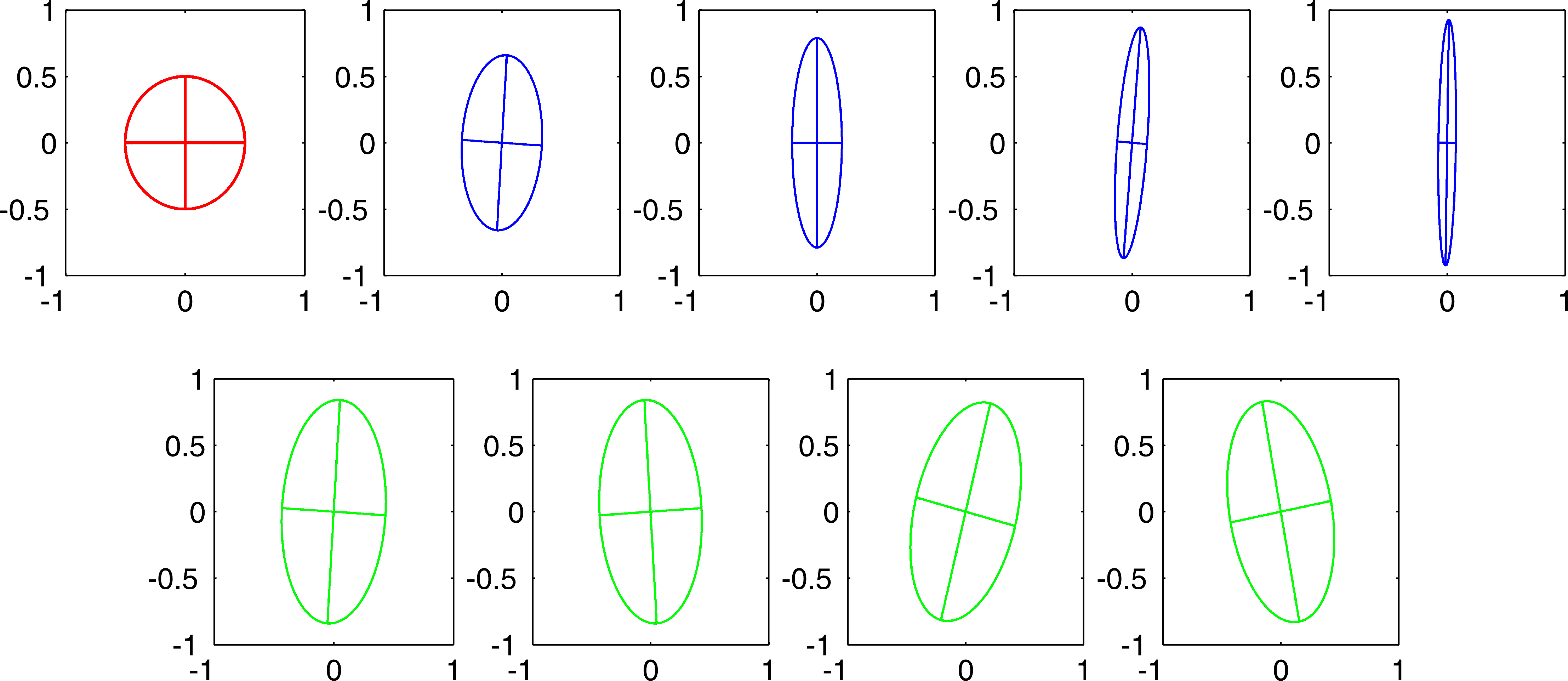}}
\caption{Sequence of Bayes updates with the new Bayes rule
\eqref{e:gbayes}: from left to right, the prior is in red; the first
data likelihood matrix is below in green; 
the first posterior is above in blue,
and so forth.}  
\label{f:ellbayes}
\end{figure}

In Figure \ref{f:ellbayes} we depict a sequence of
updates with the new Bayes rule when the data likelihood
matrix is different in each iteration. 
Observe that based on the relative lengths of the axes
(eigenvalues) and the directions of the axes (eigenvectors)
in the ellipse describing the 
current data likelihood matrix, the posterior
adjusts its axis lenghts and directions.

Other Bayes rules for our calculus are listed below. They all express one conditional in terms of the 
corresponding reverse conditional.
\begin{enumerate}[label=BR\arabic*.,ref=BR\arabic*]
\item\label{br:big}
$\D(\BS|\AS) = (\I_{\AS}\otimes \D(\BS)) \odot \D(\AS|\BS) 
\odot \left(\D(\AS)\otimes\I_{\BS}\right)^{-1},
\text{ where }
\D(\AS)=
\tr_{\BS}\bigl( (\I_{\AS}\otimes \D(\BS)) \odot \D(\AS|\BS) \bigr).$

\item\label{br:other}
$\D(\b|\AS) = \D(\b)\D(\AS|\b)\odot\D(\AS)^{-1}, \text{ where }
\D(\AS) \overset{\ref{tp:big}}{=} \tr_{\BS}\bigl((\I_{\AS}\otimes\D(\BS))\odot\D(\AS|\BS)\bigr)$.

\item\label{br:bayes}
$\D(\BS|\a) = \displaystyle\frac{\D(\BS)\odot\D(\a|\BS)}{\D(\a)},
\text{ where } \D(\a) \overset{\ref{TP:bayes}}{=} \tr(\D(\BS)\odot\D(\a|\BS))$.\\
This is the Bayes rule derived in \cite{\qbayes}
that was discussed above.

\item\label{br:small}
$\D(\b|\a) = \displaystyle\frac{\D(\b)\D(\a|\b)}{\D(\a)},
\text{ where } \D(\a) \overset{\ref{tp:small}}{=} \sum_i \D(\b_i)\D(\a|\b_i)$.\\
The summation in the normalization factor proceeds over any
orthogonal system $\b_i$.
\end{enumerate}
All these Bayes rules can be easily derived as follows:
first express the conditional on the left
i.t.o. the joint by applying the
definitions of conditional probability from Section
\ref{s:cond}; then apply these definitions again for
expressing the joint in terms of the reverse conditional.
For example,
$$\D(\BS|\a) 
\overset{\ref{cp:A.b}}{=} 
\frac{\D(\BS,\a)}{\D(\a)}
\overset{\ref{cp:a.B}}{=} 
\frac{\D(\BS)\odot\D(\a|\BS)}{\D(\a)}.
$$
As was mentioned above, the new Bayes rule can be seen as a soft maximum eigenvalue calculation.
We will now give an example that shows that its impossible to track the maximum eigenvalue without changing
the eigensystem. First, suppose that we have a diagonal density matrix $\W = \sum_i \omega_i\;\e_i\e_i^\top$
and another diagonal matrix $\S = \sum_i \sigma_i\;\e_i\e_i^\top$. 
Then $\tr(\W\S) = \sum_i \omega_i\sigma_i$ and this
means that by changing $\omega_i$ we can easily focus on the high
$\sigma_i$. Now suppose $\W$ is diagonal as before, but $\S$ has the Hadamard matrix eigensystem. 
Hadamard matrices $\H$ are square $n \times n$ matrices that have $\pm 1$ elements 
and satisfy the condition $\H\H^\top = n\I$. Thus $\frac{\H}{\sqrt{n}}$ is an orthogonal matrix.
Let $\h_i$ be the columns of this orthogonal matrix derived from a Hadamard matrix and
let $\S = \sum_i \sigma_i\; \h_i\h_i^\top$. Entries of $\h_i$ are $\pm\frac{1}{\sqrt{n}}$, 
therefore $\tr(\e_i\e_i^\top\h_j\h_j^\top) = \frac{1}{n}$. Computing the trace we obtain:
\[\tr(\W\S) = \sum_{i,j} \sigma_i\tau_j \tr(\e_i\e_i^\top\h_j\h_j^\top) = \frac{1}{n}\tr(\W)\tr(\S)
= \frac{\tr(\S)}{n}.\]
This means that 
\textit{any} diagonal density matrix $\W$ only ``sees'' the average of eigenvalues of $\S$ and 
is unable to focus on the highest eigenvalue.

\subsection{Deriving the Conventional and Generalized Bayes Rule}
In this section we show how to derive the conventional Bayes rule 
\eqref{e:bayes}
and the generalized Bayes rule for density matrices
\eqref{e:gbayes} by minimizing a tradeoff
between a relative entropy and an expected log
likelelihood. For two probability vectors $\x$ and $\y$,
the relative entropy is defined as
$\Delta(\x,\y):=\sum_i x_i \log \frac{x_i}{y_i}$.
We use the convention that $0\log 0:=0$
which is justified by $\lim_{x\rightarrow 0} x\log x=0$. 
It is well known that $\Delta(\x,\y)\geq 0$ and
that $\Delta(\x,\y) = 0$ iff $\x=\y$.

\begin{theorem}
\label{t:bayes}
Let the prior $\left(P(M_i)\right)$ be any probability vector
and the data likelihood $\left(P(y|M_i)\right)$ be any non-negative vector
of the same dimension.
Then 
$$
- \log P(y)
= \displaystyle \inf_{\left(\omega_i\right) \text{ prob.vec.}}
\quad\quad
\Delta\bigl((\omega_i),(P(M_i))\bigr)
-\sum_i \omega_i \log P(y|M_i),
$$
and $\ww=\left(P(M_i)P(y|M_i)/P(y)\right)$
is the unique optimum solution.
\end{theorem}
\begin{proof}
Let the support of a vector $\x$ be the set of all indices
$1\leq i\leq n$
s.t. $x_i\ne 0$ and denote this set as $\supp(\x)$.
For any probability vector $(\omega_i)$, such that 
$\supp((\omega_i)) \subseteq \supp(P(M_i)) \cap \supp(P(y|M_i))$,
we have
$$-\log P(y) = 
\underbrace{\sum_i \omega_i \log \frac{\omega_i}{P(M_i)}}
_{\Delta\left(\left(\omega_i\right),\left(P(M_i)\right)\right)}
-\sum_i \omega_i P(y|M_i)
-\underbrace{\sum_i \omega_i \log
\frac{\omega_i}{P(M_i)P(y|M_i)/P(y)}}
_{\Delta\left(\left(\omega_i\right),\left(P(M_i)P(y|M_i)/P(y)\right)\right)}.
$$
The precondition on the support of $(\omega_i)$ assures that
all three sums above are finite because it avoids 
the case $\omega_i \log 0$, when $\omega_i>0$.
Since the l.h.s. is a constant,
\begin{eqnarray*}
&&\displaystyle \inf_{\begin{array}{c}
(\omega_i) \text{ prob.vec.}
\\\supp((\omega_i)) \subseteq \supp(P(M_i)) \cap \supp(P(y|M_i))
\end{array}}
\Delta\left(\left(\omega_i\right),\left(P(M_i)\right)\right)
-\sum_i \omega_i P(y|M_i)
\\&=&
\displaystyle \sup_{\begin{array}{c}
(\omega_i) \text{ prob.vec.}
\\\supp((\omega_i)) \subseteq \supp(P(M_i)) \cap \supp(P(y|M_i))
\end{array}}
-\Delta\left(\left(\omega_i\right),\left(P(M_i)P(y|M_i)/P(y)\right)\right).
\end{eqnarray*}
The sup clearly has $\ww=\left(P(M_i)P(y|M_i)/P(y)\right)$
as its unique solution and
the $\inf$ remains unchanged if
the condition on the support of $(\omega_i)$ is dropped.
This gives us the statement of the theorem.
\end{proof}
This theorem can also be proven using differentiation 
(see e.g. \cite{zellner98,\expgrad,\myspeech}).
For the density matrix case this was done in
\citep{\qbayes,\meg}. We now prove the corresponding
theorem for density matrices in a different way.
For two density matrices $\A$ and $\B$,
the quantum relative entropy is defined as
$\Delta(\A,\B):=\tr(\A (\logm \A - \logm \B))$.
There is a potential problem when some of the eigenvalues of the
matrices are zero. However, we will now reason that 
this definition is justified under the assumption $0\log 0=0$ 
and $\Delta(\A,\B)$ is bounded iff $\range(\A)\subseteq \range(\B)$.

The first term $\tr(\A \logm \A)$ becomes
$\sum_i \alpha_i \log \alpha_i$, where the $\alpha_i$
are the eigenvalues of $\A$. This term is always finite.
If $\B$ is eigendecomposed as
$\sum_i \beta_i\;\b_i\b_i^\top$, then 
the second term $\tr(\A\logm\B)$ can be rewritten as 
$\sum_i \b_i^\top\A\b_i \;\log\beta_i
$. 
If $\range(\A)\subseteq\range(\B)$, then 
$\range(\B)^\perp\subseteq\range(\A)^\perp$, 
where $\perp$ denotes the orthogonal complement space. If 
$\beta_i = 0$, then $\b_i \in \range(\B)^\perp$ and under our 
assumption on $\range(\A)$ this also means that
$\b_i^\top\A\b_i = 0$. Therefore, for all $i$, s.t. $\beta_i=0$, 
the summand $\b_i^\top\A\b_i \;\log\beta_i$
has the form $0\log 0=0$.
If on the other hand, 
$\range(\A) \nsubseteq \range(\B)$, this also means
$\range(\B)^\perp \nsubseteq \range(\A)^\perp$. The
eigenvectors $\b_i$ with zero eigenvalues form a basis for
$\range(\B)^\perp$ and therefore there exists some $\b_i$ s.t. $\b_i^\top\A\b_i \neq 0$. This gives a summand
of the form $x\log 0$, with $x \neq 0$, and this is infinite. 
Notice that this discussion also means that
\begin{equation}
\label{e:logtrace}
\tr(\A\logm\B) = 
\left\{\begin{array}{ll}
\tr(\A\logm^+\B) & \text{when }\range(\A)\subseteq\range(\B) 
\\-\infty &\text{otherwise}
      \end{array}
\right.
.
\end{equation}
As before the function $\Delta(\A,\B)$ is non-negative 
and equal zero iff both arguments agree (e.g. \citep{NieChu00}).

\begin{theorem}
\label{t:gbayes}
Let the prior $\D(\MS)$ be any density matrix
and data likelihood $\D(\y|\MS)$ be any symmetric
positive definite matrix of the same dimension.
Then 
$$ 
-\log \D(\y) = \displaystyle \inf_{\W \text{ dens.mat.}} \quad\quad
\Delta(\W,\D(\MS)) -\tr(\W \logm \D(\y|\MS)),
$$ and $\W= \frac{\D(\MS) \odot \D(\y|\MS)}{D(\y)}$
is the unique optimum solution.
\end{theorem}
\begin{proof}
For any density matrix $\W$ s.t. 
$\range(\W) \subseteq \range(\D(\MS)) \cap \range(\D(\y|\MS))$,
we have
\begin{equation*}
\begin{split}
-\log D(\y) &= 
\underbrace{\tr(\W (\logm\W
-\logm\D(\MS))}_{\Delta(\W,\D(\MS))}
-\tr(\W (\logm \D(\y|\M))\\
&-\tr(\W (\logm\W - (\logm \D(\MS) + \logm \D(\y|\MS))/\D(y))).
\end{split}
\end{equation*}
Since $\range(\W) \subseteq \range(\D(\MS)) \cap
\range(\D(\y|\MS)$, $\tr(\W \logm \D(\MS))$ and
$\tr(\W \logm \D(\y|\MS))$ are both finite.
Assuming that for any symmetric positive definite matrices
$\W$, $\A$ and $\B$
\begin{equation}
\label{e:temp}
\tr(\W (\logm \A+\logm\B)
= \tr(\W \logm (\A \odot \B)),
\text{ when }
\range(\W) \subseteq \range(\A) \cap \range(\B),
\end{equation}
the above equality would become
$$-\log D(\y) = 
\Delta(\W,\D(\MS)) -\tr(\W \logm \D(\y|\M))
-\Delta(\W,(\D(\MS)\odot\D(\y|\MS))/\D(\y)).
$$
Since the l.h.s. is a constant,
\begin{eqnarray*}
&&\displaystyle \inf_{\begin{array}{c}
\W \text{ dens.mat.}
\\\range(\W) \subseteq \range(\D(\MS)) \cap \range(\D(\y|\MS))
\end{array}}
\Delta(\W,\D(\MS)) -\tr(\W \logm \D(\y|\M))
\\&=&
\displaystyle \sup_{\begin{array}{c}
\W \text{ dens.mat.}
\\\range(\W) \subseteq \range(\D(\MS)) \cap \range(\D(\y|\MS))
\end{array}}
-\Delta(\W,\left(\D(\MS)\odot\D(\y|\MS)\right)/\D(\y))
.
\end{eqnarray*}
The sup clearly has the unique solution
$\frac{\D(\MS)\odot\D(\y|\MS)}{\D(\y)}$ and the
$\inf$ remains unchanged if
the condition on the range of $\W$ is dropped.
This gives us the statement of the theorem.

We still need to show \eqref{e:temp}.
Since $\range(\W) \subseteq \range(\A) \cap \range(\B)
\overset{\eqref{i:inters}}{=} \range(\A\odot\B)
$,
\begin{eqnarray*}
\tr(\W \logm (\A\odot\B))
&\overset{\eqref{e:logtrace}}{=}& \tr(\W \logm^+(\A\odot\B))
\\&\overset{\eqref{e:plusAB}}{=}&
\tr(\W \P_{\A\cap\B} (\logm^+\A+\logm^+\B) \P_{\A\cap\B})
\\&=&\tr(\underbrace{\P_{\A\cap\B} \W \P_{\A\cap\B}}_{\W} (\logm^+\A+\logm^+\B))
\\&\overset{\eqref{e:logtrace}}{=}& \tr(\W (\logm\A+\logm\B) ).
\end{eqnarray*}
\end{proof}
We conclude with a discussion of the relationship
between the conventional Bayes rule for probability vectors
and the generalized Bayes rule for density matrices.
Density matrices are determined by a probability vector of eigenvalues 
as well as an orthogonal eigensystem. 
An orthogonal system $\w_i$ turns the prior density matrix 
$\D(\MS)$ into the probability vector 
$\left(\tr(\D(\MS) \;\w_i\w_i^\top) \right)$, 
which we call a {\em pinching} of $\D(\MS)$. 
Similarly the pinching of the data likelihood matrix
$\D(\y|\MS)$ is the vector 
$\left(\tr(\D(\y|\MS)\;\w_i\w_i^\top)\right) \in [0,1]^n$.
The idea is to express our Bayes rule for density matrices 
as the conventional Bayes rule for the pinched priors and likelihoods 
w.r.t. a certain eigensystem. That is, we want
to be able to say that the generalized Bayes rule is 
the conventional Bayes rule for the ``best'' pinching.

The above outline is essentially true, but we need to pinch
in the log domain. 
With Equality \eqref{e:logtrace}, 
Property \ref{i:rpinch} can be extended to
\begin{equation}
\label{e:rpinch}
\tr(\A \odot \u\u^\top)= e^{\u^\top \logm\A\; \u},
\text{ for {\em any} unit $\u$ and symmetric positive
definite matrix $\A$.}
\end{equation}
We call $\tr(\A \odot \w_i\w_i^\top)$
a {\em remote pinching} of $\A$.
Since its components satisfy $\tr(\D(\MS) \odot \w_i\w_i^\top)
\overset{\ref{i:ubound}}{\leq}
\tr(\D(\MS)\w_i\w_i^\top)$, the remote pinchings
of $\D(\MS)$ must be normalized to form a probability vector.

We can rewrite the argument of the 
optimization problem for the generalized Bayes rule 
based on the eigendecomposition $\WW\ww\WW^\top$
of the density matrix $\W$:
\begin{eqnarray*}
\lefteqn{\Delta(\W,\D(\MS)) -\tr(\W \logm \D(\y|\MS))}
\\&=&
\tr(\ww \WW^\top(\logm\W - \logm \D(\MS))\WW)
- \tr(\ww\WW^\top (\logm \D(\y|\MS))\WW)
\\&=& \sum_i \omega_i
(\log \omega_i - \w_i^\top (\logm \D(\MS) )\w_i)
-\sum_i \omega_i \w_i^\top (\logm \D(\y|\MS))\w_i
\\&\overset{\eqref{e:rpinch}}{=}&
\sum_i\omega_i(\log \omega_i-\log
\tr(\D(\MS)\odot\w_i\w_i^\top))
-\sum_i \omega_i \log \tr(\D(\y|\MS)\odot\w_i\w_i^\top)
\\&=&
\Delta\left(\left(\ww_i\right),\left(\tr(\D(\MS)\odot\w_i\w_i^\top)/Z_{\WW}\right)\right)
-\sum_i \omega_i \log
\tr(\D(\y|\MS)\odot\w_i\w_i^\top)-\log Z_{\WW},
\end{eqnarray*}
where the normalization $Z_{\WW}=\sum_j \tr(\D(\MS)\odot\w_j\w_j^\top)$
does not depend on the eigenvalues.
By Theorem \ref{t:bayes}, the above is minimized w.r.t.
$\ww$ when
$\ww= \left(P_\WW(M_i) P_\WW(y|M_i) / P_\WW (y)\right)$,
where $P_\WW(M_i):= \tr(\D(\MS)\odot\w_i\w_i^\top)/Z_\WW$ is the
normalized remote pinching of the prior and
$P_\WW(y|M_i):= \tr(\D(\y|\MS)\odot\w_i\w_i^\top)$ is
the remote pinching of the data likelihood matrix.
With this optimum choice of $\ww$, the minimization problem
of the generalized Bayes rule simplifies to
\begin{align*}
&\displaystyle \inf_{\WW\WW^\top=\I}\;\;\;-\log P_\WW(y)-\log Z_{\WW}\\
&\quad=\displaystyle \inf_{\WW\WW^\top=\I}\;\;\;
-\log (\sum_i \tr(\D(\MS)\odot \w_i\w_i^\top) 
\;\tr(\D(\y|\MS)\odot\w_i\w_i^\top)))
\\\quad&\overset{\ref{i:prodpinch}}{=}\displaystyle \inf_{\WW\WW^\top=\I}\;\;\;
-\log (\sum_i \tr((\D(\MS)\odot \D(\y|\MS)\odot\w_i\w_i^\top))
\\\quad&\overset{\ref{i:ubound}}{\geq}\displaystyle \inf_{\WW\WW^\top=\I}\;\;\;
-\log (\sum_i \tr((\D(\MS)\odot \D(\y|\MS)\w_i\w_i^\top))
\\\quad&=\quad \quad-\log\tr(\D(\MS) \odot \D(\y|\MS)).
\end{align*}
The above inequality is tight 
iff $\WW$ is an eigensystem of $\D(\MS) \odot \D(\y|\MS)$.
We conclude that the optimization problem for the
generalized Bayes rule is optimized when $\WW$ is
an eigensystem of $\D(\MS) \odot \D(\y|\MS)$ and the vector of eigenvalues $\ww$
is conventional posterior derived from the normalized remote pinchings
of the prior and the remote pinchings of the data likelihood.

\subsection{Chaining of the Bayes Rule}
\label{s:chain}
The conventional Bayes rule can be applied iteratively
to a sequence of data and various 
cancellations occur. For the sake of simplicity we only
consider two data points $y_1,\;y_2$:
\begin{eqnarray*}
P(M_i|y_2,y_1)
=
\frac{P(M_i|y_1) P(y_2|M_i,y_1)}
{P(y_2|y_1)}
=
\frac{P(M_i)P(y_1|M_i)P(y_2|M_i,y_1)}
{P(y_2|y_1)P(y_1)}.
\end{eqnarray*}

The normalization can be rewritten as:
\begin{eqnarray}
\nonumber
P(y_2|y_1)P(y_1)
&=&(\sum_i 
\underbrace{P(M_i|y_1)}_{\text{using } \eqref{e:bayes}} P(y_2|M_i,y_1))\;
(\sum_i P(M_i) P(y_1|M_i))
\\&=&
\sum_i P(M_i) P(y_1|M_i) P(y_2|M_i,y_1) 
= P(y_2,y_1). 
\label{e:prod}
\end{eqnarray}

Analogously, by essentially applying the generalized Bayes rule
\eqref{e:gbayes} two times we get:
\begin{eqnarray*}
\D(\MS|\y_2,\y_1)
=\frac{\D(\MS|\y_1)\odot\D(\y_2|\MS,\y_1) }
{\D(\y_2|\y_1)}
= \frac{\D(\MS)\odot\D(\y_1|\MS)\odot\D(\y_2|\MS,\y_1)}
{\D(\y_2|\y_1)\D(\y_1)}
.
\end{eqnarray*}
As in the diagonal case \eqref{e:prod}, the normalization
can be rewritten into one term (by applying
\ref{tp:basic} twice and then the generalized Bayes rule
\eqref{e:gbayes}):
\begin{eqnarray*}
\D(\y_2|\y_1)\D(\y_1) &=&
\tr(\D(\MS|\y_1)\odot\D(\y_2|\MS,\y_1))
\;\;\tr(\D(\MS)\odot\D(\y_1|\MS))\\
&=& \tr(\frac{\D(\MS)\odot\D(\y_1|\MS)}{\tr(\D(\MS)\odot\D(\y_1|\MS))}
\odot\D(\y_2|\MS,\y_1))
\;\;\tr(\D(\MS)\odot\D(\y_1|\MS)\\
&=& \tr(\D(\MS)\odot\D(\y_1|\MS)\odot\D(\y_2|\MS,\y_1)) = \D(\y_1,\y_2)
.
\end{eqnarray*}

Finally as in \eqref{e:expected}, 
we can upper bound the data probability $\D(\y_1,\y_2)$ 
in terms of the product of the expected variances for the two trials:
\begin{eqnarray*}
\D(\y_2,\y_1) &=& 
\tr(\D(\MS|\y_1)\odot\D(\y_2|\MS,\y_1))\;\;
\tr(\D(\MS)\odot\D(\y_1|\MS))\\
&\leq& \tr(\D(\MS|\y_1)\D(\y_2|\MS,\y_1))\;\;
\tr(\D(\MS)\D(\y_1|\MS)).
\end{eqnarray*}

\subsection{Bounds}
\label{ss:bounds}

Recall the following conventional bound for
the negative log-likelihood of the data i.t.o.
the negative log-likelihood of the MAP estimator:
\begin{equation}
\label{e:cmap}
\begin{split}
-\log P(y) &= -\log \sum_i P(y|M_i) P(M_i)
\\&\leq \min_i (-\log P(y|M_i) -\log P(M_i)).
\end{split}
\end{equation}
We will give analogous bound for density matrices.
For this we need the following inequality:
For any unit vector $\m$ and symmetric positive definite
matrix $\A$:
\begin{equation}
\label{e:logbound}
-\log \m^\top \A \m 
\overset{\ref{i:ubound}}{\leq}
-\log \tr(\A \odot \m \m^\top)
\overset{\eqref{e:rpinch}}{=} -\m^\top (\logm \A)\: \m.
\end{equation}
Using the fact that $\tr(\A) \geq \m^\top\A\m$,
we can now prove an analogous MAP bound for the generalized probabilities:
\begin{eqnarray*}
-\log \D(\y)& =& -\log \tr(\D(\y|\MS) \odot \D(\MS))
\\&\leq&\min_{\m} (-\log \m^\top(\D(\y|\MS)\odot \D(\MS))\:\m)
\\&\overset{\eqref{e:logbound}}{\leq}&\min_{\m} (-\m^\top \logm (\D(\y|\MS)\odot\D(\MS))\:\m)
\\&\leq& \min_{\m} (-\m^\top\logm \D(\y|\MS)\:\m -\m^\top\logm\D(\MS)\:\m).
\end{eqnarray*}
The last inequality becomes Equality \eqref{e:temp}, 
when $\m\in \range(\D(\MS))\cap\range(\D(\y|\MS)).$
Otherwise, it holds trivially because
$-\m^\top \logm (\D(\y|\MS)\odot\D(\MS))\:\m=+\infty$.

Intuitively, there are two domains: the probability domain
and the log probability domain. The conventional bound
\eqref{e:cmap} can also be written in the probability
domain:
$$P(y) \geq \max_i P(M_i) P(y|M_i).$$
However for the generalized probability case, there
does not seem to be a simple similar inequality
in the probability domain.
Throughout the paper we always notice that the matrix operations
need to be done in the log domain.

In the conventional case $P(y)$ is also upper bounded by
$\max_i P(y|M_i)$. For the generalized case,
the analogous formula is the following, where
$\mu_i$ and $\m_i$ are the eigenvalues/vectors of
$\D(\MS)$ and $\m$ any unit direction:
\begin{eqnarray*}
\D(\y) &=&
\tr(\D(\y|\MS)\odot\D(\MS))
\\&\leq& \tr(\D(\y|\MS)\D(\MS))
\\&=& \sum_i \mu_i\; \m_i^\top \D(\y|\MS)\:\m_i
\\&\leq& \max_i\; \m_i^\top \D(\y|\MS)\:\m_i\\
&\leq& \max_{\m}\; \m^\top \D(\y|\MS)\m.
\end{eqnarray*}

\section{Summary of the Probability Calculus for Density Matrices}
\label{s:formulas}
In this section we give a summary of 
all the rules of our calculus.
The definitions are indicated with $:=$
and at the end we summarize the justification
for our choice of definitions.
Table \ref{t:chart} shows 
connections between different objects and the formulas that 
relate them.
\begin{table}
\begin{tabular}{cc}
\begin{minipage}{0.5\textwidth}
\centerline{\includegraphics[width=0.2\textwidth]{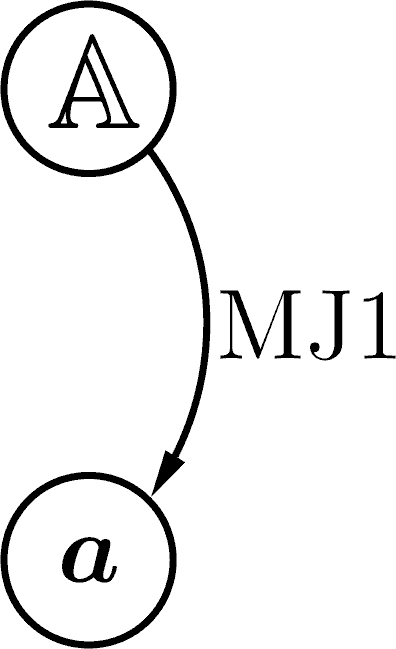}}
\end{minipage}&
\begin{minipage}{0.5\textwidth}
\centerline{\includegraphics[width=0.4\textwidth]{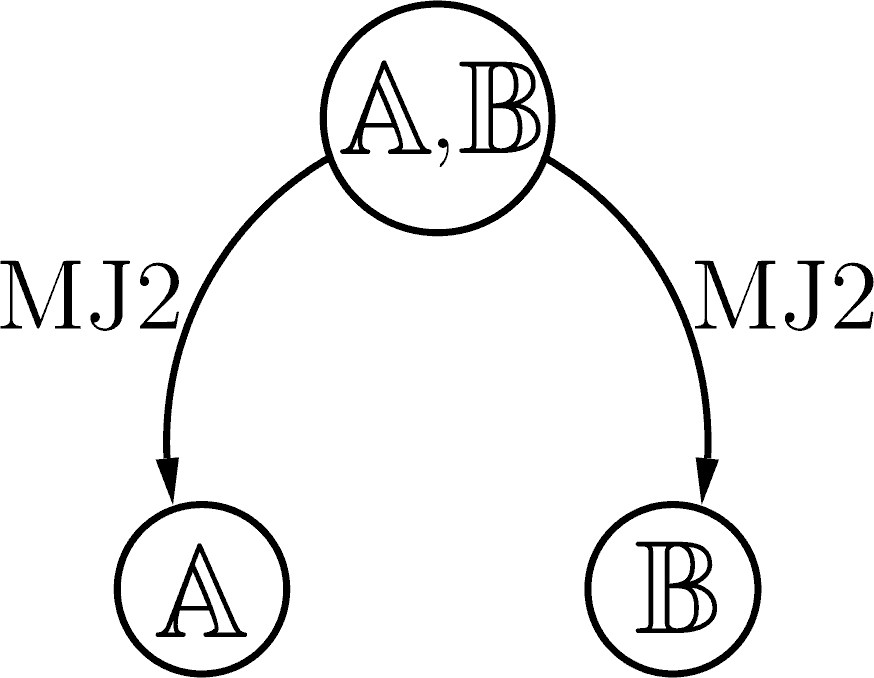}}
\end{minipage}\\
\begin{minipage}{0.5\textwidth}
\centerline{\includegraphics[width=0.6\textwidth]{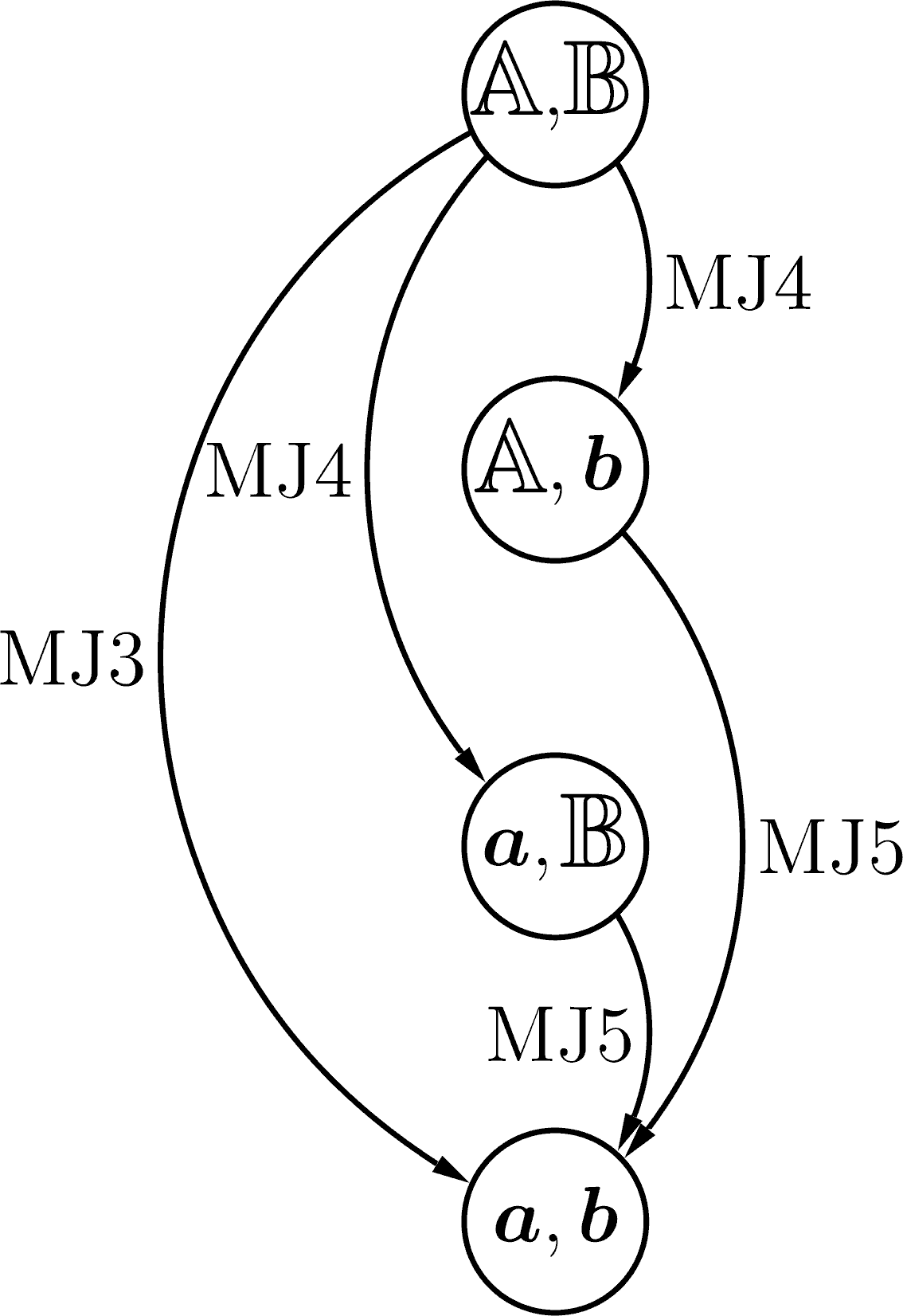}}
\end{minipage}&
\begin{minipage}{0.5\textwidth}
\centerline{\includegraphics[width=0.6\textwidth]{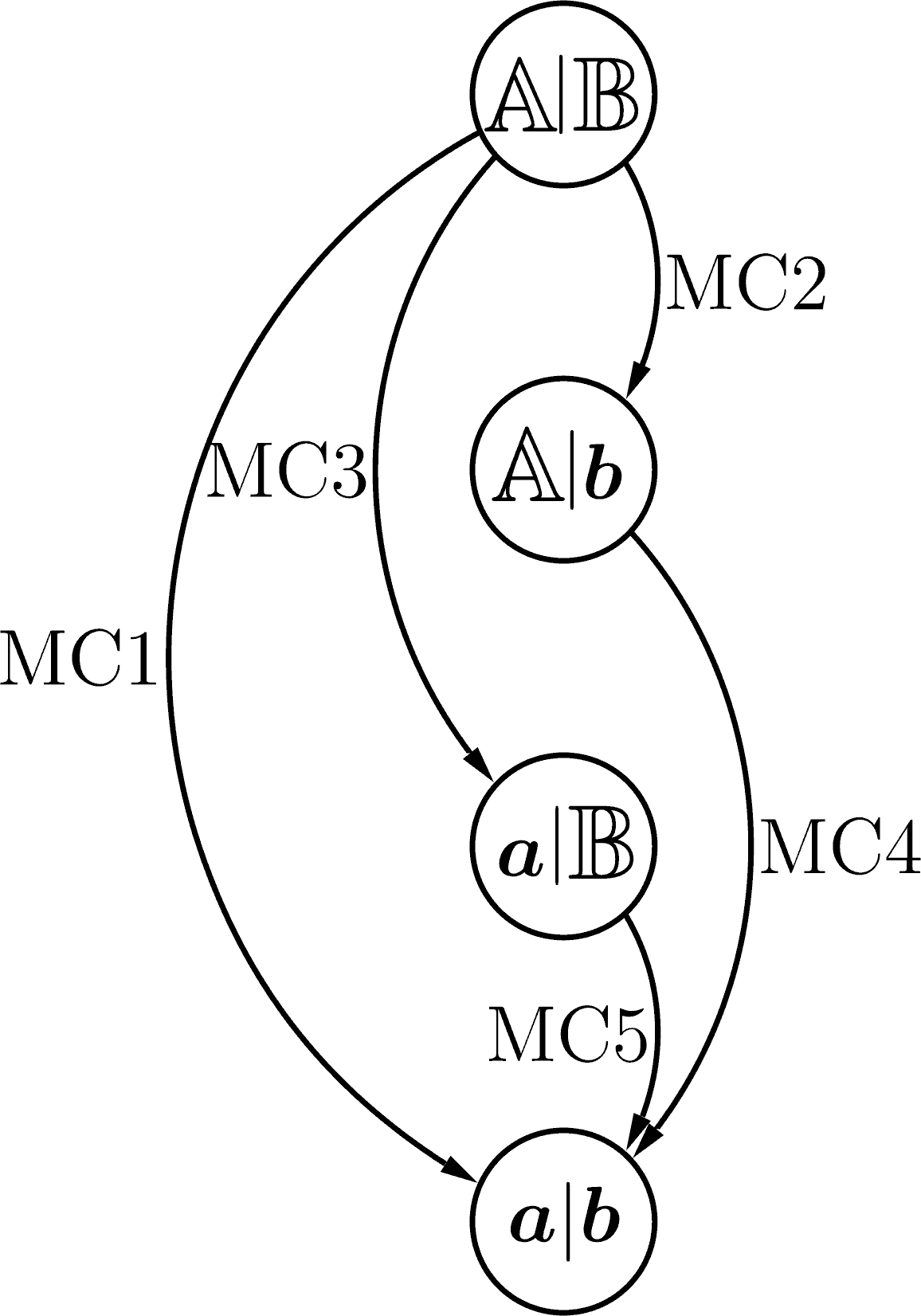}}
\end{minipage}\\
\multicolumn{2}{c}{%
\begin{minipage}{1\textwidth}
\centerline{\includegraphics[width=0.4\textwidth]{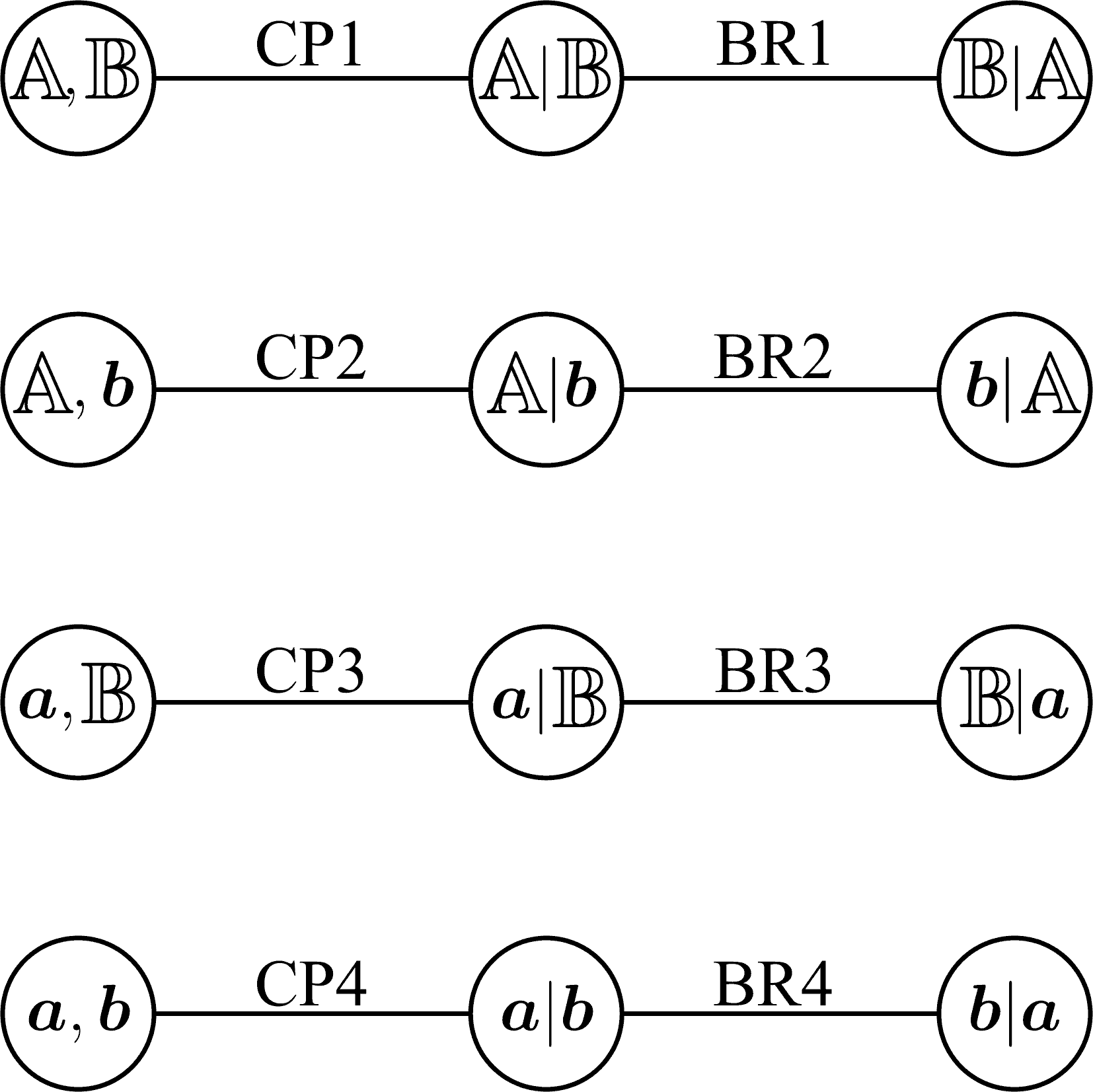}}
\end{minipage}
}
\end{tabular}
\caption{A series of charts summarizing the different relationships for joints and conditionals. Each
edge references the formula stating the relationship. For symmetric cases only one formula is given
and the corresponding edges in the chart will have the same label.}
\label{t:chart}
\end{table}

\subsection{Marginalization Rules for Joints of Sections
\ref{s:joint} and \ref{s:marginal}}

\begin{enumerate}[label=MJ\arabic*.,ref=MJ\arabic*]
\item $\D(\a) := \tr(\D(\AS)\a\a^{\top}) = \a^{\top}\D(\AS)\a$.
\item $\D(\AS) := \tr_\BS(\D(\AS,\BS))$.
\item $\D(\a,\b) := \tr(\D(\AS,\BS)(\a\otimes\b)(\a\otimes\b)^{\top}) = 
\tr(\D(\AS,\BS)(\a\a^{\top} \otimes \b\b^{\top}))$.
\item $\D(\AS,\b) := \tr_\BS(\D(\AS,\BS)(\I_\AS\otimes\b\b^{\top}))$.
\item $\D(\a,\b) = \tr(\D(\AS,\b)\a\a^{\top})$.
\end{enumerate}

\subsection{Conditional Probability Rules of Section \ref{s:cond}}
\begin{enumerate}[label=CP\arabic*.,ref=CP\arabic*]
\item $\D(\AS|\BS) := \D(\AS,\BS)\odot(\I_\AS\otimes\D(\BS))^{-1}$.

\item $\displaystyle\D(\AS|\b) := \frac{\D(\AS,\b)}{\tr(\D(\AS,\b))}$

\item $\D(\a|\BS) := \D(\a,\BS)\odot\D(\BS)^{-1}$

\item $\displaystyle\D(\a|\b) := \frac{\D(\a,\b)}{\D(\b)}$
\end{enumerate}

\ref{cp:A.B} has the form:
density matrix $\odot$ inverse of a normalization.
Below we reexpress the other definitions in this unified form:

\begin{enumerate}[label=CP$'$\arabic*.,ref=CP$'$\arabic*,start=2]
\item $\displaystyle\D(\AS|\b) = \tr_\BS(\D(\AS,\BS)(\I_\AS\otimes\b\b^{\top}))
\odot\tr_\BS((\I_\AS\otimes\D(\BS))(\I_\AS\otimes\b\b^{\top}))^{-1}$.

\item $\D(\a|\BS) = \tr_\AS(\D(\AS,\BS)(\a\a^{\top}\otimes\I_\BS))
\odot\tr_\AS((\I_\AS\otimes\D(\BS))(\a\a^{\top}\otimes\I_\BS))^{-1}$.

\item $\displaystyle\D(\a|\b) = \tr(\D(\AS,\BS)(\a\a^{\top}\otimes\b\b^{\top}))
\odot
\tr((\I_\AS\otimes\D(\BS))(\a\a^{\top}\otimes\b\b^{\top}))^{-1}$.
\end{enumerate}

\subsection{Marginalization Rules for Conditionals of
Section \ref{s:cond}}
\begin{enumerate}[label=MC\arabic*.,ref=MC\arabic*]
\item $\D(\a|\b) = \displaystyle\frac{\tr(\D(\AS|\BS)\odot
(\I_\AS\otimes\D(\BS))(\a\a{\top}\otimes\b\b^{\top}))}
{\tr(\D(\BS)\b\b^{\top})}$
\item $\D(\AS|\b) = \displaystyle\frac{\tr_\BS(\D(\AS|\BS)\odot(\I_\AS\otimes\D(\BS))
(\I_\AS\otimes\b\b^{\top}))}
{\tr(\D(\BS)\b\b^{\top})}$
\item $\D(\a|\BS) = \tr_\AS(\D(\AS|\BS)\odot(\I_\AS\otimes\D(\BS))(\a\a^{\top}\otimes\I_\BS))
\odot \D(\BS)^{-1}$.
\item\label{ii:ok} $\D(\a|\b)=\tr(\D(\AS|\b)\a\a^{\top})$.
\item $\D(\a|\b) = \displaystyle\frac{\tr((\D(\a|\BS)\odot\D(\BS))\,\b\b^{\top})}{\tr(\D(\BS)\b\b^{\top})}$.
\end{enumerate}
All the rules here except for \ref{ii:ok}
require additional information for marginalization,
which was not necessary in the conventional case. See discussion of marginalization of conditionals
in Section \ref{s:cond}.

\subsection{Theorems of Total Probability of Section \ref{s:total}}
\begin{enumerate}[label=TP\arabic*.,ref=TP\arabic*]
\item $\D(\a) = \sum_i \D(\a|\b_i)\D(\b_i)$ for any
orthogonal system $\b_i$ of space $\BS$.

\item $\D(\AS) = \sum_i \D(\AS, \b_i)$ 
for any 
orthogonal system $\b_i$ of space $\BS$.

\item $\D(\AS) = \tr_\BS(\D(\AS|\BS)\odot(\I_\AS\otimes\D(\BS)))$.

\end{enumerate}

\subsection{Bayes Rules of Section \ref{s:bayes}}
\begin{enumerate}[label=BR\arabic*.,ref=BR\arabic*]
\item $\D(\BS|\AS) = (\I_{\AS}\otimes \D(\BS)) \odot \D(\AS|\BS) 
\odot \left(\D(\AS)\otimes\I_{\BS}\right)^{-1},
\text{ where }
\D(\AS)=
\tr_{\BS}\bigl( (\I_{\AS}\otimes \D(\BS)) \odot \D(\AS|\BS) \bigr).$

\item $\D(\a|\BS) = \D(\a)\D(\BS|\a)\odot\D(\BS)^{-1},
\text{ where } \D(\BS) = \tr_\AS(\D(\BS|\AS)\odot(\D(\AS)\otimes\I_\BS))$.

\item $\D(\BS|\a) = \displaystyle\frac{\D(\BS)\odot\D(\a|\BS)}{\D(\a)},
\text{ where } \D(\a) = \tr(\D(\BS)\odot\D(\a|\BS))$.

\item $\D(\b|\a) = \displaystyle\frac{\D(\a|\b)\D(\b)}{\D(\a)},
\text{ where } \D(\a) = \sum_i \D(\a|\b_i)\D(\b_i)$ and the
summation is over any 
orthogonal system $\b_i$.
\end{enumerate}

\subsection{Summary of Justifications for the Definitions}
Note that only the rules MJ1-4 and CP rules are definitions. Everything else in our calculus can be derived
from these. MJ1 is justified by Gleason's Theorem as
discussed in Section \ref{s:gleason}. 
Gleason's Theorem also justifies
MJ3, where the Kronecker product provides 
the natural way to specify a joint unit 
(See discussion in Section \ref{s:joint}). 
MJ2 is standard in quantum physics 
and $\tr_{\BS} (\D(\AS,\BS))$ was shown to be a density
matrix in Lemma \ref{l:partdens}. The rule is also compatible 
with the conventional case as well as with the
natural generalization of independence discussed in
Section \ref{s:marginal}.
MJ4 is the natural definition of $\D(\AS,\b)$ that satisfies MJ5
and is compatible with the conventional case.

We will outline how CP2 can be motivated as a quantum 
relative entropy projection.
For positive definite matrices $\A$ and $\B$, we extend the definition of quantum relative entropy
as follows: 
$\Delta(\A,\B)=\tr(\A (\logm \A-\logm\B) +\B -\A)$.
Note that this ``unnormalized'' relative entropy, 
coincides with the standard one when $\A$ and $\B$ have
trace one.
Now CP2 is motivated as
$$\D(\AS|\b)=\arginf_{\W \text{ dens. mat.}} 
\quad\quad
\Delta\left(\W,\D(\AS,\b)\right).$$
CP3 is motivated analogous to the generalized Bayes rule 
(See Section \ref{s:bayes}):
$$\D(\a,\B)=
\quad\quad
\arginf_{\W}\; \Delta(\W,\D(\BS)) - \tr(\W\logm\D(\a|\BS)).
$$
CP1 can be motivated in a similar fashion, but now
the variable is over the joint space $(\AS,\BS)$: 
$$\D(\A,\B)=
\quad\quad
\arginf_{\W \text{ dens.mat.}}\; \Delta(\W,\I_{\AS} \otimes \D(\BS)) 
- \tr(\W\logm\D(\AS|\BS)).
$$
CP1 also was previously used in \cite{ca-qecp-99}
to allow a suitable definition of conditional quantum
entropy.
Finally, the last rule CP4 was chosen in analogy to the
conventional case. It also has an interpretation
as two successive  quantum measurements (see Appendix \ref{a:condmeas}).

Historically, we first justified the generalized Bayes rule \ref{br:bayes}
based on the minimum relative entropy principle
(See \cite{\qbayes} and Section \ref{s:bayes}). After that
we chose definitions \ref{cp:A.B}-\ref{cp:a.b} 
to be compatible with this generalized Bayes rule.

\section{Conclusions}
Density matrices are central to quantum physics.
We utilize many mathematical techniques from that field
to develop a Bayesian probability calculus for density matrices. 
Intuitively, the new calculus will be useful when
the data likelihood $\D(\y|\MS)$ has non-zero off-diagonal
elements, i.e. information about which components are correlated
or anti-correlated.
The main new operation $\A\odot\B$ first takes logs of the 
matrices adds the logs and finally exponentiates. 
Any straightforward implementation of the $\odot$ operation
requires the eigendecompositions of the matrices,
which are expensive to obtain. Throughout our work we notice that
the log domain seems to be more important in the matrix case.

Interestingly enough the $\odot$ operation
has also been employed in computer graphics for combining
affine transformation \cite{alexa}.
Also the simulation of quantum computations
based on the Lie Trotter Formula
(\cite{NieChu00}, Chapter 4.7)
can be interpreted as applying the $\odot$ operation to unitary
matrices and not to symmetric positive definite matrices
as we do in this paper.

The main update in quantum physic is a unitary evolution
of the current density matrix $\A$, i.e.
$\A:=\U \A \U^\top,$ where $\U$ is unitary.
For example, the main differential equation for 
density matrices in quantum physics
is the following version of the Schr\"odinger Equation
\cite{feynman}:
$$\frac{\partial \D(\MS|t)}{\partial t}
=i\: (\H\:\D(\MS|t)-\D(\MS|t)\:\H),
\text{ where $\H$ is skew Hermitian.}
$$
The solution has the form
$$\D(\MS|t)=\expm(-i\:t\:\H)\: \D(\MS|0)\: \expm(i\:t\:\H),$$
where $\D(\MS|0)$ is the initial density matrix.
Since $i\:t\:\H$ is skew Hermitian, both
exponentials are unitary. Thus the above update represents 
a unitary transformation of the initial density matrix
$\D(\MS|0)$.
Such transformations leave the eigenvalues unchanged
and only affect the eigensystem.
In contrast our generalized Bayes rule updates
both the eigenvalues and eigenvectors, and the 
conventional Bayes rule can be seen as only updating
the eigenvalues while keeping the eigenvectors fixed.
Therefore the Bayes rules are decidedly not unitary updates.

For the sake of completeness we now express the Bayes rules
also as solutions to differential equations.
In the conventional case, the differential equations are
($1\leq i\leq n$):
$$ \frac{\partial \log P(M_i|t)}{\partial t} 
= \log P(y|M_i) - \sum_j P(M_j|t) \log P(y|M_j).$$
The solution is 
$$ P(M_i|t) = \frac{P(M_i|0)  P(y|M_i)^t}
                   {\sum_j P(M_j|0)  P(y|M_j)^t}.$$
If we take the value $P(M_i|0)$ as the prior $P(M_i)$ 
then the expression for $P(M_i|1)$ 
becomes the conventional Bayes rule
\eqref{e:bayes}.
There is a similar differential equation for
the generalized Bayes rule
(For the sake of simplicity we assume that
the prior $\D(\MS)$
and data likelihood matrix $\D(\y|\MS)$ are strictly positive definite):
$$ \frac{\partial \logm \D(\MS|t)}{\partial t} 
= \logm\D(\y|\MS) - \tr(\D(\MS|t) \logm \D(\y|\MS)).$$
The solution has the form 
\begin{eqnarray*} 
\D(\MS|t) &=& 
\frac{\expm\left(\logm \D(\MS|0) + t \logm \D(\y|\MS)\right)}
{\tr\left(\expm\left(\logm \D(\MS|0) + t \logm \D(\y|\MS)\right)\right)}
\\&\stackrel{\eqref{e:explog}}{=}&
\frac{\D(\MS|0) \odot \D(\y|\MS)^t}
{\tr\left( \D(\MS|0) \odot \D(\y|\MS)^t\right)}.
\end{eqnarray*}
If we set $\D(\MS|0)$ to the prior $\D(\MS)$,
then the expression for $\D(\MS|1)$ becomes the generalized Bayes rule
\eqref{e:gbayes}.
Notice again that the differential equations emphasize
the log domain and that the $\odot$ operation appears in
the solution. 

At this point we have no convincing application for
the new probability calculus. However, a similar
methodology was used to derive and prove bounds for 
parameter updates of density matrices that led
to a version of Boosting \cite{\meg} where
the distribution over the examples 
is replaced by a density matrix,
an online variance minimization algorithm
where the parameter space is the unit ball \cite{\variance},
and an on-line algorithm for Principal Component Analysis
\cite{\pca}. 

In this paper our parameters expressing
the uncertainty are symmetric positive definite matrices.
However using essentially the EG$\pm$ transformation \citep{\expgrad},
it has been shown recently that 
inference can be done with arbitrarily 
shaped matrices \cite{\subspace}.
This leaves the strong possibility that the calculus developed here will
generalize to arbitrary shaped matrices as well.
In that case the elementary events are ``asymmetric dyads''
$\u\v^\top$ and the underlying decomposition is the
SVD decomposition. 

The new calculus seems to be rich enough to bring
out some of the interesting phenomena of
quantum physics, such as superposition and entanglement.
Maybe the new calculus can be used to maintain
``uncertainty'' in quantum computation.

On a more technical note, we conjecture that
for all non-decoupled joints $\D(\AS,\BS)$ there
is a one-to-one mapping to the conditionals 
$\D(\AS|\BS)$, and the EM-like algorithm given
in Section \ref{s:cond} converges to $\D(\BS)$, s.t.
$\D(\AS,\BS)=\D(\AS|\BS)\odot (\I_\AS\otimes\D(\BS))$.

Finally, we will reason in a simple case that
generalized probability space is more ``connected''
and a clever algorithm might be able to exploit this.
Assume zero is encoded as the distribution $(1,0)$
and one as the distribution $(0,1)$. 
Moving from the zero distribution to the one distributions
can be done by lowering the probability of the first
component and increasing the probability of the second.
As density matrices, zero and one would be
$\left(\begin{smallmatrix}
1&0\\
0&0
\end{smallmatrix}
\right)
$
and
$
\left(
\begin{smallmatrix}
0&0\\
0&1
\end{smallmatrix}
\right)
$, respectively.
Note that the eigensystem for both matrices
is the identity matrix and 
there is now a second way to go from zero to one 
that keeps the eigenvalues/probabilities fixed
but swaps the eigenvectors:
\[
\begin{pmatrix}
0&1\\
1&0
\end{pmatrix}
\begin{pmatrix}
1&0\\
0&0
\end{pmatrix}
\begin{pmatrix}
0&1\\
1&0
\end{pmatrix}
=
\begin{pmatrix}
0&0\\
0&1
\end{pmatrix}
.\]
\section*{Acknowledgment}
Many thanks to Torsten Ehrhardt who first proved
to us the range intersection property \ref{i:inters} 
and the $\logm^+$ formula \ref{i:logplus}
for the $\odot$ operation.

\bibliographystyle{alpha}
\bibliography{bibs/coltlocal,bibs/colt}

\newpage
\appendix
\noindent{\bf \Large APPENDIX}

\section{Quantum-Mechanical Interpretation of Conditional Probability $\D(\a|\b)$}
\label{a:condmeas}
We will now show how to interpret the conditional probability 
$\D(\a|\b)$ in terms of two quantum measurements. 
The two measurements will be performed one after another
on the joint density $\D(\AS,\BS)$
and $\D(\a|\b)$ will be a probability of outcome 1
for the second measurement given the first measurement had outcome 1. 
First, we measure $\D(\AS,\BS)$ with event $\I_{\AS}\otimes\b\b^\top$. 
Assume that we get outcome 1. 
Using the generalization of collapse rule for events 
(see e.g. \cite{NieChu00}), 
the successor density matrix can be computed as follows:
\[\widehat{\D}(\AS,\BS) = 
\frac{(\I_{\AS}\otimes\b\b^\top)\D(\AS,\BS)(\I_{\AS}\otimes\b\b^\top)}
{\tr((\I_{\AS}\otimes\b\b^\top)\D(\AS,\BS)(\I_{\AS}\otimes\b\b^\top))} \]
The second measurement consists of measuring the updated joint 
with event $\a\a^\top\otimes\I_\BS$.
Now the probability for getting outcome 1 is computed as:
\begin{eqnarray*}
\tr(\widehat{\D}(\AS,\BS)(\a\a^\top\otimes\I_\BS))
&=& \frac{\tr((\I_{\AS}\otimes\b\b^\top)\D(\AS,\BS)(\I_{\AS}\otimes\b\b^\top)(\a\a^\top\otimes\I_\BS))}
{\tr((\I_{\AS}\otimes\b\b^\top)\D(\AS,\BS)(\I_{\AS}\otimes\b\b^\top))}\\
&\overset{\ref{kp:prod}+\text{cycle}\;\;}{=}&
\frac{\tr(\D(\AS,\BS)(\a\a^\top\otimes\b\b^\top))}
{\tr(\D(\AS,\BS)(\I_{\AS}\otimes\b\b^\top))} = \frac{\D(\a,\b)}{\tr(\D(\AS,\BS)(\I_{\AS}\otimes\b\b^\top))}
.
\end{eqnarray*}
The denominator can be simplified 
using partial trace properties:
\[\tr(\D(\AS,\BS)(\I_{\AS}\otimes\b\b^\top)) \overset{\ref{pt:2}}{=}
\tr(\tr_{\AS}(\D(\AS,\BS)(\I_{\AS}\otimes\b\b^\top))) \overset{\ref{pt:3}}{=}
\tr(\underbrace{\tr_{\AS}(\D(\AS,\BS))}_{\D(\BS)}\b\b^\top) = \D(\b).
\]
Therefore the probability of outcome 1 
on the second measurement (given the first outcome was 1) is: 
\[\tr(\widehat{\D}(\AS,\BS)(\a\a^\top\otimes\I_\BS)) 
= \frac{\D(\a,\b)}{\D(\b)} \overset{\ref{cp:a.b}}{=} \D(\a|\b).\]
\end{document}